\RequirePackage[l2tabu, orthodox]{nag}
\documentclass[11pt]{article}

\usepackage[T1]{fontenc}

\usepackage[bitstream-charter]{mathdesign}
\usepackage{amsmath}
\usepackage[scaled=0.92]{PTSans}

\usepackage[
  paper  = letterpaper,
  left   = 1.65in,
  right  = 1.65in,
  top    = 1.0in,
  bottom = 1.0in,
  ]{geometry}

\usepackage[usenames,dvipsnames]{xcolor}
\definecolor{shadecolor}{gray}{0.9}

\usepackage[final,expansion=alltext]{microtype}
\usepackage[english]{babel}
\usepackage[parfill]{parskip}
\usepackage{afterpage}
\usepackage{framed}

\usepackage{lineno}

\usepackage{ragged2e}

\newcounter{parcount}

\usepackage{booktabs}
\usepackage{multirow}
\usepackage{tabularx}

\usepackage{listings}
\usepackage{fancyvrb}
\fvset{fontsize=\normalsize}

\usepackage{natbib}

\usepackage[colorlinks,linktoc=all]{hyperref}
\usepackage[all]{hypcap}
\hypersetup{citecolor=BurntOrange}
\hypersetup{linkcolor=black}
\hypersetup{urlcolor=MidnightBlue}

\usepackage[acronym,nowarn]{glossaries}

\lstdefinestyle{mystyle}{
    commentstyle=\color{OliveGreen},
    keywordstyle=\color{BurntOrange},
    numberstyle=\tiny\color{black!60},
    stringstyle=\color{MidnightBlue},
    basicstyle=\ttfamily,
    breakatwhitespace=false,
    breaklines=true,
    captionpos=b,
    keepspaces=true,
    numbers=left,
    numbersep=5pt,
    showspaces=false,
    showstringspaces=false,
    showtabs=false,
    tabsize=2
}
\usepackage[colorinlistoftodos,
           prependcaption,
           textsize=small,
           backgroundcolor=yellow,
           linecolor=lightgray,
           bordercolor=lightgray]{todonotes}

\usepackage{amsthm}  

\usepackage{bm}
\usepackage[Export]{adjustbox}

\usepackage[colorinlistoftodos,
           prependcaption,
           textsize=small,
           backgroundcolor=yellow,
           linecolor=lightgray,
           bordercolor=lightgray]{todonotes}

\usepackage{graphicx}
\usepackage{caption}
\usepackage{subcaption}
\usepackage{wrapfig}

\usepackage{booktabs}
\usepackage{arydshln} \usepackage{multirow}

\usepackage{listings}
\usepackage{fancyvrb}
\fvset{fontsize=\normalsize}

\hypersetup{linkcolor=black}
\hypersetup{urlcolor=BurntOrange}

\usepackage{listings}
\lstdefinestyle{alp_style}{
    commentstyle=\color{OliveGreen},
    numberstyle=\tiny\color{black!60},
    stringstyle=\color{BrickRed},
    basicstyle=\ttfamily\scriptsize,
    breakatwhitespace=false,
    breaklines=true,
    captionpos=b,
    keepspaces=true,
    numbers=none,
    numbersep=5pt,
    showspaces=false,
    showstringspaces=false,
    showtabs=false,
    tabsize=2
}

\usepackage{capt-of}

\usepackage{lipsum}

\usepackage{authblk}
\usepackage{enumitem}

\theoremstyle{remark}
\newtheorem*{lemma*}{Lemma}

\usepackage{algorithm}
\usepackage{algpseudocode}
\usepackage{rotating}

\def\eqref#1{equation~\ref{#1}}

\def\1{\bm{1}}

\DeclareMathAlphabet{\mathsfit}{\encodingdefault}{\sfdefault}{m}{sl}
\SetMathAlphabet{\mathsfit}{bold}{\encodingdefault}{\sfdefault}{bx}{n}

\usepackage{authblk}
\usepackage{enumitem} 

\usepackage{textcomp}

\usepackage{multirow}

\usepackage{subcaption}
\usepackage{graphicx}

\usepackage{multicol}
\usepackage{adjustbox}
\usepackage{booktabs}
\usepackage{longtable}

\usepackage{bbding}
\usepackage{pifont}

\usepackage{listings}
\definecolor{codegreen}{rgb}{0,0.6,0}
\definecolor{codegray}{rgb}{0.5,0.5,0.5}
\definecolor{codepurple}{rgb}{0.58,0,0.82}
\definecolor{backcolour}{rgb}{0.95,0.95,0.92}
\lstdefinestyle{mystyle}{
    backgroundcolor=\color{backcolour},   
    commentstyle=\color{codegreen},
    keywordstyle=\color{magenta},
    numberstyle=\tiny\color{codegray},
    stringstyle=\color{codepurple},
    basicstyle=\ttfamily\footnotesize,
    breakatwhitespace=true,         
    breaklines=true,                 
    captionpos=b,                    
    keepspaces=true,                                  
    showspaces=false,                
    showstringspaces=false,
    showtabs=false,                  
    tabsize=2
}
\usepackage{placeins}

\usepackage{amsmath}

\usepackage{mwe}
\usepackage{makecell}
\usepackage{titletoc}

\title{\textbf{LLM4Mat-Bench: Benchmarking Large Language Models for Materials Property Prediction}}

\author[1,2]{Andre Niyongabo Rubungo}
\author[3]{Kangming Li}
\author[3,4,5,6]{Jason Hattrick-Simpers}
\author[1,2,*]{\\Adji Bousso Dieng}
\affil[1]{Department of Computer Science, Princeton University}
\affil[2]{\href{https://vertaix.princeton.edu/}{Vertaix}}
\affil[3]{Acceleration Consortium, University of Toronto}
\affil[4]{Department of Materials Science and Engineering, University of Toronto}
\affil[5]{Vector Institute for Artificial Intelligence}
\affil[6]{Schwartz Reisman Institute for Technology and Society}
\affil[*]{Corresponding author: \texttt{adji@princeton.edu}}

\begin{document}
\maketitle

\begin{abstract}
\noindent Large language models (LLMs) are increasingly being used in materials science. However, little attention has been given to benchmarking and standardized evaluation for LLM-based materials property prediction, which hinders progress. We present LLM4Mat-Bench, the largest benchmark to date for evaluating the performance of LLMs in predicting the properties of crystalline materials. LLM4Mat-Bench contains about 1.9M crystal structures in total, collected from 10 publicly available materials data sources, and 45 distinct properties. LLM4Mat-Bench features different input modalities: crystal composition, CIF, and crystal text description, with 4.7M, 615.5M, and 3.1B tokens in total for each modality, respectively. We use LLM4Mat-Bench to fine-tune models with different sizes, including LLM-Prop and MatBERT, and provide zero-shot and few-shot prompts to evaluate the property prediction capabilities of LLM-chat-like models, including Llama, Gemma, and Mistral. The results highlight the challenges of general-purpose LLMs in materials science and the need for task-specific predictive models and task-specific instruction-tuned LLMs in materials property prediction \footnote{The Benchmark and code can be found at: \url{https://github.com/vertaix/LLM4Mat-Bench}}.\\

\end{abstract}

\section{Introduction} \label{sec:intro}
With the remarkable success of large language models (LLMs) in solving natural language tasks \citep{devlin2018bert,radford2018improving,raffel2020exploring,achiam2023gpt,touvron2023llama} and different scientific tasks \citep{lin2022language,edwards2022translation,valentini2023promises,castro2023large,fang2023mol,lv2024prollama}, scientists have recently started to leverage LLMs to tackle very important and challenging problems in materials science, including predicting materials properties \citep{rubungo2023llm, korolev2023accurate, xie2023darwin,das2023crysmmnet, qu2024leveraging, choudhary2024atomgpt, li2024probing} and discovering new materials \citep{antunes2023crystal,gruver2023fine, qu2024leveraging, choudhary2024atomgpt}. 

The learning capabilities of LLMs have the potential to revolutionize the field of materials science. For example, recent research by \citet{rubungo2023llm} has demonstrated the exceptional performance of LLMs in predicting the properties of crystalline materials based on textual descriptions of their structures. In their study, they introduced a novel dataset, TextEdge, which comprises textual descriptions of crystals and their corresponding properties. This dataset was used to fine-tune the encoder component of the T5-small model for the task of materials property prediction. The findings of \citet{rubungo2023llm} challenge the conventional practice of heavily relying on graph neural networks and using solely either crystal composition or structure as input for property prediction. Their work underscores the significance of further investigating the extent to which LLMs can be harnessed to develop innovative techniques for accurately predicting the properties of crystalline materials, thereby enhancing the materials discovery pipeline. Unfortunately, the proposed TextEdge dataset is limited in scope, comprising approximately 145K samples and encompassing only three distinct properties. Furthermore, its lack of diversity, being derived from a single data source (Materials Project \citep{jainmaterials}), hinders its effectiveness in assessing the robustness of LLMs in materials property prediction.

In this work, we introduce LLM4Mat-Bench, a benchmark dataset collected to evaluate the performance of LLMs in predicting the properties of crystalline materials. To the best of our knowledge, LLM4Mat-Bench is the most extensive benchmark to date for assessing the efficacy of language models in materials property prediction. The dataset comprises approximately two million samples, sourced from ten publicly available materials sources, each containing between 10K and 1M structure samples. LLM4Mat-Bench encompasses several tasks, including the prediction of electronic, elastic, and thermodynamic properties based on a material's composition, crystal information file (CIF), or textual description of its structure. We use LLM4Mat-Bench to evaluate several LLMs of different sizes, namely LLM-Prop \citep{rubungo2023llm} (35M parameters), MatBERT \citep{walker2021impact} (109.5M parameters), and Llama variants \citep{touvron2023llama, dubey2024llama} (3B, 7B, and 8B parameters), Mistral variants \citep{jiang2023mistral} (7B parameters), and Gemma variants \citep{team2024gemma, team2024gemmaB} (7B and 9B parameters). And we provide fixed train-valid-test splits, along with carefully designed zero-shot and few-shot prompts to ensure reproducibility. We anticipate that LLM4Mat-Bench will significantly advance the application of LLMs in addressing critical challenges in materials science, including property prediction and materials discovery.

\section{LLM4Mat-Bench}

\subsection{Data Collection Process} \label{sec:data_collection}
We collected the data used to create LLM4Mat-Bench from 10 publicly available materials data sources. In this section, we describe each data source and discuss how we accessed its data.

\subsubsection{Data sources} \label{sec-datasources}
hMOF      \citep{wilmer2012large} is a publicly available database\footnote{https://mof.tech.northwestern.edu/} consisting of about 160K Metal-Organic Frameworks (MOFs), generated by Wilmer et al. using computational approaches. Materials Project (MP)     \citep{jainmaterials} is a database with around 150K materials, offering free API access\footnote{https://next-gen.materialsproject.org/api} for data retrieval, including CIF files and material properties. The Open Quantum Materials Database (OQMD)     \citep{kirklin2015open} is a publicly accessible database\footnote{https://www.oqmd.org/} of 1.2M materials, containing DFT-calculated thermodynamic and structural properties, created at Northwestern University. OMDB     \citep{borysov2017organic} is an organic materials database with about 12K structures and related electronic band structure properties, freely available\footnote{https://omdb.mathub.io/}. JARVIS-DFT     \citep{choudhary2017high, choudhary2018computational} is a repository created by NIST researchers, containing around 75.9K material structures with downloadable properties\footnote{https://jarvis.nist.gov/jarvisdft}. QMOF     \citep{rosen2021machine, rosen2022high} is a quantum-chemical property database of over 16K MOFs, accessible via GitHub\footnote{https://github.com/Andrew-S-Rosen/QMOF}. JARVIS-QETB     \citep{garrity2023fast} is a NIST-created database\footnote{https://jarvis.nist.gov/jarvisqetb/} of nearly one million materials with tight-binding parameters for 65 elements. GNoME is a database of 381K new stable materials discovered by \citet{merchant2023scaling} using graph networks and DFT, available on GitHub\footnote{https://github.com/google-deepmind/materials\_discovery/blob/main/DATASET.md}. Cantor HEA     \citep{li2024efficient} is a DFT dataset of formation energies for 84K alloy structures, available on Zenodo\footnote{https://doi.org/10.5281/zenodo.10854500}. SNUMAT is a database with around 10K experimentally synthesized materials and DFT properties, accessible via API\footnote{https://www.snumat.com/apis}.

\subsubsection{Collecting crystal information files (CIFs) and materials property}

Crystal structure files (CIFs), material compositions, and material properties were collected from publicly accessible sources described in Section \ref{sec-datasources}. Data collection was facilitated by APIs and direct download links provided by the respective databases. For databases such as Materials Project, OMDB, SNUMAT, JARVIS-DFT, and JARVIS-QETB, user registration is required for access, while databases like hMOF, QMOF, OQMD, and GNoME allow direct data access without registration. From each source, we obtained CIFs and associated material properties. Although the Materials Project and JARVIS-DFT databases offer a broader range of properties, we selected a subset—10 and 20 properties respectively—that adequately represents the data within our benchmark, based on the number of data points available for each property. This selection was made to optimize computational efficiency when training models across the 65 properties included in LLM4Mat-Bench.

\subsubsection{Generating the textual description of crystal structure} \label{text-generation}
LLMs perform better with textual input, and \citet{rubungo2023llm, korolev2023accurate, qu2024leveraging} have demonstrated that LLMs can effectively learn the structural representation of a crystal from its textual description, outperforming graph neural network (GNN)-based models that directly utilize the crystal structure for property prediction. 

Crystal structures are typically described in file formats such as Crystallographic Information File (CIF) which include predominantly numbers describing lattice vectors and atomic coordinates and are less amenable to LLMs. Instead of directly using these as inputs, we use Robocrystallographer    \citep{jainmaterials} to deterministically generate texts that are more descriptive of crystal structures from CIF files. Robocrystallographer was developed and has been used by the Materials Project team to auto-generate texts for their database. Given a structure, Robocrystallographer leverages predefined rules and existing libraries to extract chemical and structural information, including oxidation states, global structural descriptions (symmetry information, prototype matching, structural fingerprint calculations etc.), and local structural descriptions (e.g. bonding and neighbor analysis, connectivity). 
This method not only generate deterministic and human-readable texts, but also ensures no data contamination in our fine-tuned LLMs, as the data sources mentioned do not include these crystal text descriptions.

\begin{table}[!htbp]
\centering
\caption{LLM4Mat-Bench statistics.}
\resizebox{1.0\textwidth}{!}{
\begin{tabular}{l l l l l l l l l l l l l}

\toprule 

\multirow{2}{*}{\textbf{Data source}} & \multirow{2}{*}{\textbf{\# Structure files}} & \multicolumn{4}{c}{\textbf{\# Structure-Description pairs}} & \multicolumn{3}{c}{\textbf{\# Tokens (Words)}} & \multicolumn{3}{c}{\textbf{\# Avg. subword tokens/Sample}} & \multirow{2}{*}{\textbf{\# Properties}}  \\

\cmidrule(r){3-6} \cmidrule(r){7-9} \cmidrule(r){10-12}

& & \textbf{Total} & \textbf{Train} & \textbf{Validation} & \textbf{Test} & \textbf{Composition} & \textbf{Structure} & \textbf{Description}& \textbf{Composition} & \textbf{Structure} & \textbf{Description}& \\

 \midrule
 
OQMD  \citep{kirklin2015open} & 1,008,266 & 964,403 & 771,522 & 96,440 & 96,441 & 964K & 96M & 244M & 5.3 & 635.4 & 347.3 & 2 \\
JARVIS-QETB  \citep{garrity2023fast} & 829,576 & 623,989 & 499,191 & 62,399 & 62,399 & 624K & 45M & 90M & 3.5 & 466.6 & 202.5 & 4 \\
GNoME  \citep{merchant2023scaling} & 381,000 & 376,276 & 301,020 & 37,628 & 37,628 & 830K & 78M & 508M & 9.7 & 1185.3 & 1711.3 & 6 \\
Materials Project  \citep{jainmaterials} & 146,143 & 146,143  & 125,825 & 10,000 & 10,318 & 272K & 37M & 157M & 6.8 & 1611.8 & 1467.3 & 10\\
hMOF  \citep{wilmer2012large} & 133,524 & 132,743 & 106,194 & 13,274 & 13,275 & 449K & 96M & 581M & 14.9 & 4583.9 & 5629.3 & 7 \\
Cantor HEA  \citep{li2024efficient} & 84,024 & 84,019 & 67,215 & 8,402 & 8,402 & 84K & 11M & 251M & 9.5 & 868.4 & 4988.6 & 4\\
JARVIS-DFT  \citep{choudhary2017high, choudhary2018computational} & 75,965 & 75,965 & 60,772 & 7,596 & 7,597 & 76K & 9M & 25M & 5.0 & 786.0 & 455.9 & 20 \\
QMOF  \citep{rosen2021machine, rosen2022high} & 16,340 & 7,656 & 6,124 & 766 & 766 & 8K & 7M & 22M & 14.0 & 5876.4 & 3668.0 & 4\\
OMDB  \citep{borysov2017organic} & 12,500 & 12,122 & 9,697 & 1,212 & 1,213 & 66K & 8M & 14M & 14.8 & 4097.4 & 1496.6 & 1 \\
SNUMAT \footnote{https://www.snumat.com/} & 10,481 & 10,372 & 8,297 & 1,037 & 1,038 & 16K & 2M & 4M & 5.9 & 1244.5 & 539.1 & 7\\
\midrule 
\textbf{Total} & 2,697,779 & 1,978,985 & 1,592,315 & 193,357 & 193,313 & 4.7M & 615.5M & 3.1B & 7.9 & 1559.7 & 1703.6 & 65\\

\bottomrule

\end{tabular}}

\label{llm4matbench_stats}
\end{table}

\begin{table}[!htbp]
\centering
\caption{Comparing the LLM4Mat-Bench with other existing benchmarks.}
\resizebox{1.0\textwidth}{!}{
\begin{tabular}{l c c c c c c c c c c }

\toprule 

\multirow{2}{*}{\bf Benchmark} & \multirow{2}{*}{\bf{\# Data Sources}} & \multirow{2}{*}{\bf{\# Distinct Properties}} & \multicolumn{3}{c}{\textbf{\# Properties/\# Samples}} & \multicolumn{2}{c}{\textbf{\# Properties/Task Type}} & \multicolumn{3}{c}{\textbf{Material Representations}} \\

 \cmidrule(r){4-6} \cmidrule(r){7-8}  \cmidrule(){9-11}

& & & \bf <10k & \bf 10-100k & \bf 100k+ & \bf Regression & \bf Classification & \bf Composition & \bf Structure & \bf Description\\ 

\midrule

MatBench  \citep{dunn2020benchmarking} & 6 & 10 & 7 & 3 & 3 & 10 & 3 & \ding{51} & \ding{51} & \ding{55} \\
TextEdge  \citep{rubungo2023llm} & 1 & 3 & 0 & 0 & 3 & 2 & 1 & \ding{55} & \ding{55} & \ding{51}\\
LLM4Mat-Bench (Ours) & 10 & 45 & 5 & 31 & 29 & 60 & 5 & \ding{51} & \ding{51} & \ding{51}\\ 

\bottomrule

\end{tabular}}
\label{tab:benchmarks_comparison}
\end{table}

\subsection{Data Statistics}
As Table \ref{llm4matbench_stats} shows, LLM4Mat-Bench comprises 2,697,779 structure files, which, after pairing with descriptions generated by Robocrystallographer and filtering out descriptions with fewer than five words, result in 1,978,985 composition-structure-description pairs\footnote{The total number of pairs were 2,433,688, after removing about 454703 duplicated pairs across datasets, it resulted to 1,978,985 pairs.}. The reduction in sample count is also due to Robocrystallographer's inability to describe certain CIF files. The total samples for each dataset in LLM4Mat-Bench are randomly split into 80\%, 10\%, and 10\% for training, validation, and testing, respectively. OQMD has the highest number of samples at 964,403, while QMOF has the fewest with 7,656 samples. On average, each dataset in LLM4Mat-Bench contains approximately 200,000 samples.

In LLM4Mat-Bench, when combined, textual descriptions contain 3.1 billion tokens, crystal structures 615 million, and compositions 4.7 million\footnote{We used NLTK toolkit as a tokenizer to count the number of words/tokens.}. OQMD leads in composition tokens (964K), while hMOF has the most description tokens (581M). For CIFs, both OQMD and hMOF have around 96M tokens. On average, compositions have 8 subword tokens per sample, CIFs 1600, and descriptions 1700. hMOF averages the longest inputs for compositions (14.9) and descriptions (5629), while QMOF leads in structures (5876.4)\footnote{We used Llama 2 tokenizer to count the number of subword tokens.}. JARVIS-DFT has the most tasks with 20 properties, followed by Materials Project with 10, and OMDB with one. Details on sample counts are in Section \ref{sec:discussion}.

LLM4Mat-Bench provides the most comprehensive dataset compared to existing benchmarks, with the largest number of samples, properties, and tasks, including 60 regression and 5 classification tasks (see Table \ref{tab:benchmarks_comparison}). It also offers more diverse material representations, incorporating chemical formulas, crystal structures, and crystal text descriptions. In contrast, MatBench  \citep{dunn2020benchmarking} and TextEdge  \citep{rubungo2023llm} have fewer tasks and less representation diversity, with MatBench lacking crystal text descriptions and TextEdge missing material compositions and crystal structures.

\subsection{Data Quality}
Since Robocrystallographer generates crystal text descriptions in a deterministic manner following predefined and well-validated rules  \citep{jainmaterials}, these texts should faithfully describe the crystal structures used to generate them. Regarding the quality of labels, they are calculated from simulations and are usually considered noise-free. Properties data except those from JARVIS-QETB and hMOF are obtained from DFT, which is based on fundamental quantum mechanical equations. While DFT calculations can still be performed with different levels of approximations and fidelity, the DFT-calculated properties are usually considered to be highly reliable and are routinely used as noise-free ground truths for ML models in the materials science community.

\section{Results}

\subsection{Experimental Details}\label{sec:experiments}
We conducted about 1,235 experiments, evaluating the performance of five models and three material representations on each property for each data source. Consistent with standard practices in materials science, we evaluated performance separately for each data source rather than combining samples from different sources for the same property. This approach accounts for variations in techniques and settings used by different data sources, which can result in discrepancies, such as differing band gaps for the same material. Below, we will describe each material representation, model, and metric that we used to conduct our experiments.

\subsubsection{Material Representations}
LLM4Mat-Bench includes three distinct materials representations: Composition, CIF, and Description (see Table \ref{tab:llm4matbench_input_formats}). The primary goal of using these diverse representations is to identify which best enhances LLM performance in predicting material properties across different data sources.

\textbf{Composition (Comp.)} 
Material composition refers to the chemical formula of a material. Though it only provides stoichiometric information, studies have shown it can still be a reliable material representation for property prediction   \citep{dunn2020benchmarking, tian2022information}. For LLMs, it offers the advantage of being a short sequence that usually fits within the model's context window, making it efficient to train. To further optimize efficiency, we set the longest sequence of material compositions from each data source as the context window, rather than using the default 512 tokens for fine-tuning while the original length is kept during inference.

\textbf{CIF} We represent the materials structure using CIF files, the conventional way of representing the crystal structure in crystallography   \citep{hall1991crystallographic}. CIFs are commonly used for GNN-based models, but some recent works have demonstrated that it can also work with LLMs   \citep{antunes2023crystal,flam2023language,  gruver2023fine}. 

\textbf{Description (Descr.)} As we outlined in Section \ref{text-generation}, we also use textual descriptions of crystal structures as representations for both atomic crystals and MOFs.

\subsubsection{Models} \label{sec:models_info}
We benchmarked different LLM-based models with various sizes and types, and a GNN-based baseline. Herein, We provide the details of each model.

\textbf{CGCNN}   \citep{xie2018crystal} is employed as a GNN baseline which is widely used in the materials science community\footnote{Although CGCNN is not state-of-the-art for some properties, it was faster compared to models like ALIGNN   \citep{choudhary2021atomistic} and DeeperGatGNN   \citep{omee2022scalable}, making it suitable for our extensive experiments}. We trained it on LLM4Mat-Bench from scratch with optimal hyperparameters: 128 hidden dimensions, batch size of 256, three message passing layers, 1e-2 learning rate, 8.0 radius cutoff, 12 nearest neighbors, and 500 training epochs, though extending to 1000 epochs improved performance in some cases.

\textbf{MatBERT}   \citep{walker2021impact} is a BERT-base model   \citep{devlin2018bert} with 109 million parameters, pretrained on two million materials science articles. We fine-tuned MatBERT on LLM4Mat-Bench, following \citet{rubungo2023llm}, and achieved optimal performance with a 512-token input length, 64-sample batch size, 5e-5 learning rate, 0.5 dropout, and 100 epochs using the Adam optimizer and onecycle learning rate scheduler   \citep{smith2019super}. Although training for 200 epochs improves performance, results are reported for 100 epochs due to computational constraints.

\textbf{LLM-Prop} is a model based on the encoder part of T5-small model   \citep{raffel2020exploring} introduced by \citet{rubungo2023llm}, with 35 million parameters, smaller than MatBERT. It predicts material properties from the textual descriptions of crystal structures. To adapt LLM-Prop on CIF, we employed xVal encoding   \citep{golkar2023xval}, where we parse an input sequence $x$ to extract numerical values into a list $x_{\text{num}}$, replace them with a [NUM] token to form $x_{\text{text}}$, and then embed $x_{\text{text}}$, followed by multiplying each [NUM] embedding by its corresponding value in $x_{\text{num}}$ to get $h_{\text{embed}}$ that we feed to the model. xVal encoding ensures that the quantitative value of each number is reflected in the input embedding while reducing the input length caused by the high volume of numerical values in CIF files, which extend the length of the input sequence after tokenization. We fine-tuned LLM-Prop on LLM4Mat-Bench and optimizing with a 1e-3 learning rate, 0.2 dropout, Adam optimizer, and onecycle learning rate scheduler for 100 epochs, with a 768-token input length, batch size of 64 for training, and 512 for inference. While \citet{rubungo2023llm} recommended that training for 200 epochs and increasing the number of input tokens improves the performance, we could not replicate this due to computational constraints.

\begin{table}[hbt!]
\caption{\label{tab:llama_prompt_templates} Prompt template. <material representation type> denotes \textit{``chemical formula"}, \textit{``cif structure"}, or \textit{``structure description"}. <value> represents the input context (for example \textit{NaCl}, etc.). <property name> denotes the name of the property (for example \textit{band gap}, etc.). <predicted value> represents the property value generated by Llama 2 while <actual value\_i> represents the ground truth of the \textcolor{olive}{EXAMPLE\_i}. \textcolor{brown}{FINAL PROMPT} and \textcolor{teal}{ RESPONSE} denote the input prompt to Llama 2 and its generated output, respectively.}
\resizebox{1.0\textwidth}{!}{
\begin{tabular}{l l}

\toprule

\bf Prompt Type & \bf Template \\

\midrule

\multirow{9}{*}{-} & \textcolor{blue}{SYSTEM PROMPT:}\\
& <<SYS>> \\
& You are a material scientist.\\
& Look at the <material representation type> of the given crystalline material and  predict its property.\\
& The output must be in a json format. For example:  \{property\_name:predicted\_property\_value\}.\\
& Answer as precise as possible and in as few words as possible.\\
& <</SYS>>\\
& \\

& \textcolor{purple}{INPUT PROMPT:} \\
& <material representation type>: <value>\\
& property name: <property name>.\\
\midrule

\multirow{2}{*}{0-shot}& \textcolor{brown}{FINAL PROMPT:} <s>[INST] + \textcolor{blue}{SYSTEM PROMPT} + \textcolor{purple}{INPUT PROMPT} +  [/INST] \\
& \textcolor{teal}{RESPONSE:} <property name>:<predicted value>\\
\midrule

\multirow{7}{*}{5-shot}& \textcolor{olive}{EXAMPLE\_1:} \qquad \qquad \qquad \qquad \qquad \qquad \qquad \qquad... \qquad \qquad \textcolor{olive}{EXAMPLE\_5:}\\
& <material representation type>: <value\_1> \qquad \qquad \qquad <material representation type>: <value\_5>\\
& property name: <property name>. \qquad \qquad \qquad \qquad \qquad property name: <property name>.\\
& <property name>:<actual value\_1> \qquad \qquad \qquad \qquad \quad \; <property name>:<actual value\_5>\\
\\

& \textcolor{brown}{FINAL PROMPT:} <s>[INST] + \textcolor{blue}{SYSTEM PROMPT} +  \textcolor{olive}{EXAMPLE\_1} + ... +  \textcolor{olive}{EXAMPLE\_5} + \textcolor{purple}{INPUT PROMPT} + [/INST] \\
& \textcolor{teal}{RESPONSE:} <property name>:<predicted value>\\

\bottomrule

\end{tabular}}
\end{table}

\begin{figure}
    \centering
    \includegraphics[width=\linewidth]{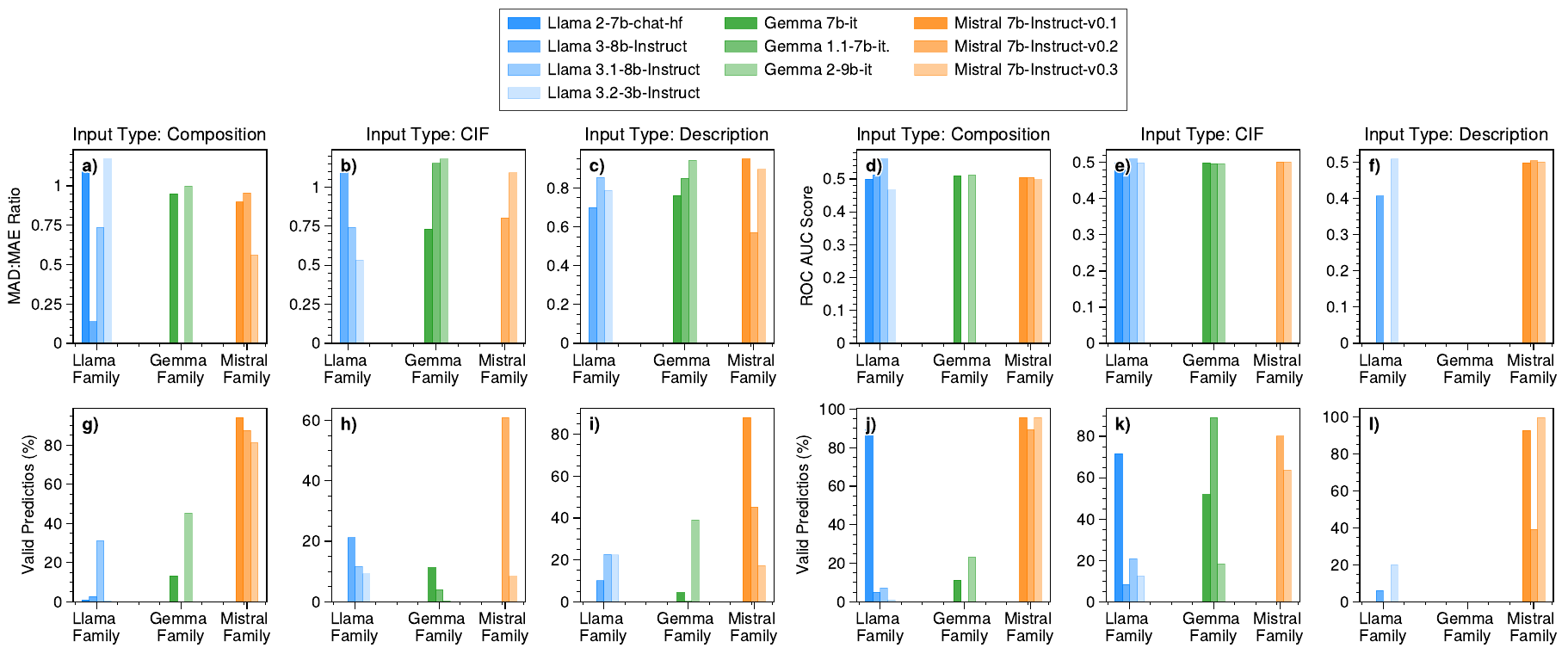}
    \caption{The performance comparison of different chat-based LLM versions is presented with results based on 5-shot prompts, averaged over three inference runs. Panels (a)–(c) and (d)–(f) show each model's accuracy in predicting band gaps and stability in the MP dataset, respectively, while panels (g)–(i) and (j)–(l) depict the percentage of valid predictions for band gap and stability on the test set.}
    \label{fig:chat-llms-comparison}
\end{figure}

\textbf{Llama, Gemma, and Mistral} To assess the performance of conversational LLMs in materials property prediction, we tested the currently available variants of Llama, Gemma, and Mistral using our designed zero-shot and five-shot prompts (see Table \ref{tab:llama_prompt_templates}) without fine-tuning. For the CIF structure prompts, we removed "\textit{\# generated using pymatgen}" comment that is appended to each file. The maximum input length was set to the maximum number of tokens each model can handle while the output length was set to 256, with a batch size of 256 samples, temperature of 0.8, and top-K sampling applied with $K=10$. The inference details of each model can be found in Appendix \ref{sec:inference_details}. For five-shot examples, we sampled from crystals with shorter structures and descriptions to reduce the context length. We also made sure the property values for those examples are diverse (for instance, they should not all have 0.0 eV as their bandgap values). For each model family, we first compared all the associated models for band gap and stability prediction on the MP test set and found no significant performance gain from one variant to another (see Figure \ref{fig:chat-llms-comparison}). Therefore, for the remaining properties in each dataset, we conducted experiments using only one variant from each family as its representative. Additionally, we reported both zero-shot and five-shot performance for Llama, while focusing solely on five-shot performance for other models due to the significant performance gap observed between zero-shot and five-shot scenarios.

We trained all models using NVIDIA RTX A6000 GPUs. Training MatBERT with two GPUs on about 300K data points and 100 epochs took about four days while for LLM-Prop took about 2.5 days. For CGCNN, it took about 7 hours training time on one GPU for 500 epochs. With one GPU, Llama 2 took about a half day to generate the output of 40K samples with 256 tokens maximum length each. We report the test set results averaged over five runs for predictive models and three runs for generative models.

\subsubsection{Evaluation Metrics}
Following \citet{choudhary2021atomistic}, we evaluated regression tasks using the ratio between the mean absolute deviation (MAD) of the ground truth and the mean absolute error (MAE) of the predicted properties. The MAD:MAE ratio ensures an unbiased model comparison between different properties where the higher ratio the better. According to \citet{choudhary2021atomistic}, a good predictive model should have at least 5.0 ratio. MAD values represent the performance of a random guessing model predicting the average value for each data point. To provide a comprehensive performance comparison across datasets, we also reported the weighted average of MAD:MAE across all properties in each dataset (Wtd. Avg. (MAD:MAE), see Equation \ref{wamade_eq}).

 For classification tasks, we reported the area under the ROC curve (AUC) for each task and provided the weighted average across all properties (Wtd. Avg. AUC, see Equation \ref{waauc_eq}).

\begin{equation} \label{wamade_eq}
    \text{Wtd.\, Avg.\,(MAD/MAE)} = \frac{\sum_{\text{i}}^{m} \text{TestSize}_{\text{i}} \times \frac{\text{MAD}_{\text{i}}}{\text{MAE}_{\text{i}}}}{\sum_{\text{i}}^{m} \text{TestSize}_{\text{i}}},
\end{equation}

\begin{equation} \label{waauc_eq}
    \text{Wtd.\, Avg.\,AUC}=\frac{\sum_{\text{i}}^{m} \text{TestSize}_{\text{i}}\times \text{AUC}_{\text{i}}}{\sum_{\text{i}}^{m} \text{TestSize}_{\text{i}}},
\end{equation}

$m$ denotes the number of regression properties in the dataset.

\begin{table}[!htbp]
\caption{The Wtd. Avg. (MAD:MAE) scores (the higher the better) for the regression tasks in the LLM4Mat-Bench are reported. \textbf{Bolded} results indicate the best model for each input format, while \colorbox[RGB]{204, 229, 255}{bolded results with blue background} highlight the best model per each dataset. \colorbox[RGB]{255, 204, 229}{Inval.} denotes cases where the Llama model failed to generate outputs with a property value or had fewer than 10 valid predictions.}

\centering

\resizebox{1.0\textwidth}{!}{
\begin{tabular}{l l c c c c c c c c c c }

\toprule

\multirow{3}{*}{\textbf{Input} }& \multirow{3}{*}{\textbf{Model}} & \multicolumn{10}{c}{\bf Dataset}\\

\cmidrule(r){3-12}
 
&  & \textbf{MP} & \textbf{JARVIS-DFT} & \textbf{GNoME} & \textbf{hMOF} & \textbf{Cantor HEA} & \textbf{JARVIS-QETB} & \textbf{OQMD} & \textbf{QMOF} & \textbf{SNUMAT} & \textbf{OMDB}\\

& & 8 tasks & 20 tasks & 6 tasks & 7 tasks & 4 tasks & 4 tasks & 2 tasks & 4 tasks & 4 tasks & 1 task\\

 \midrule

 CIF & CGCNN (baseline) & 5.319 & 7.048 & 19.478 & 2.257 & 17.780 & 61.729 & 14.496 & 3.076 & 1.973 & 2.751 \\
 
 \midrule

 \multirow{6}{*}{Comp.} & Llama 2-7b-chat:0S &  0.389  & \colorbox[RGB]{255, 204, 229}{Inval.}  & 0.164  & 0.174 & 0.034 & 0.188 & 0.105 & 0.303 & 0.940 & 0.885  \\
 
 & Llama 2-7b-chat:5S & 0.627 & 0.704 & 0.499 & 0.655 & 0.867 & 1.047 & 1.160 & 0.932 & 1.157 & 1.009\\

 & Mistral 7b-Instruct-v0.1:5S & 0.505 & 0.313 & 0.336 & 0.375 & 0.23 & 0.643 & 0.601 & 0.545 & 0.703 & 0.222\\

 & Gemma 2-9b-it:5S & 0.758 & 0.49 & 0.495 & 0.624 & 0.508 & 0.718 & 0.647 & 0.925 & 0.843 & 0.967\\
  
 & MatBERT-109M & \textbf{5.317} & \textbf{4.103} & 12.834 & 1.430 & 6.769 & 11.952  & 5.772 & \bf 2.049 & \bf 1.828 & \textbf{1.554} \\
 
 & LLM-Prop-35M & 4.394 & 2.912 & \bf 15.599  & \bf 1.479 & \bf 8.400 & \bf 59.443 & \bf 6.020 & 1.958 & 1.509 & 1.507 \\
 
 \midrule

\multirow{6}{*}{CIF} & Llama 2-7b-chat:0S & 0.392 & 0.216 & 6.746 & 0.214 & 0.022 & 0.278 & 0.028 & 0.119 & 0.682 & 0.159 \\

& Llama 2-7b-chat:5S & \colorbox[RGB]{255, 204, 229}{Inval.} & \colorbox[RGB]{255, 204, 229}{Inval.} & \colorbox[RGB]{255, 204, 229}{Inval.} & \colorbox[RGB]{255, 204, 229}{Inval.} & \colorbox[RGB]{255, 204, 229}{Inval.} & 1.152 & 1.391 & \colorbox[RGB]{255, 204, 229}{Inval.} & \colorbox[RGB]{255, 204, 229}{Inval.} & 0.930 \\

& Mistral 7b-Instruct-v0.1:5S & \colorbox[RGB]{255, 204, 229}{Inval.} & 0.328 & 0.174 & \colorbox[RGB]{255, 204, 229}{Inval.} & 0.095 & 0.251 & 0.236 & \colorbox[RGB]{255, 204, 229}{Inval.} & \colorbox[RGB]{255, 204, 229}{Inval.} & \colorbox[RGB]{255, 204, 229}{Inval.}\\

& Gemma 2-9b-it:5S & \colorbox[RGB]{255, 204, 229}{Inval.} & \colorbox[RGB]{255, 204, 229}{Inval.} & 0.642 & \colorbox[RGB]{255, 204, 229}{Inval.} & 0.8 & 0.763 & 0.957 & \colorbox[RGB]{255, 204, 229}{Inval.} & \colorbox[RGB]{255, 204, 229}{Inval.} & \colorbox[RGB]{255, 204, 229}{Inval.}\\
 
& MatBERT-109M & 7.452 & 6.211 & 14.227 & 1.514 & 9.958 & 47.687 & 10.521 & 3.024 & \bf 2.131 & \textbf{1.777} \\

& LLM-Prop-35M & \textbf{8.554} & \textbf{6.756} & \textbf{16.032} & \bf 1.623 & \bf 15.728 & \colorbox[RGB]{204, 229, 255}{\textbf{97.919}} & \colorbox[RGB]{204, 229, 255}{\textbf{11.041}} & \colorbox[RGB]{204, 229, 255}{\textbf{3.076}} & 1.829 & \bf 1.777\\

\midrule

\multirow{6}{*}{Descr.} & Llama 2-7b-chat:0S & 0.437 & 0.247 & 0.336 & 0.193 & 0.069 & 0.264 & 0.106 & 0.152 & 0.883 & 0.155 \\

& Llama 2-7b-chat:5S & 0.635 & 0.703 & 0.470 & 0.653 & 0.820 & 0.980 & 1.230 & 0.946 & 1.040 & 1.001 \\

& Mistral 7b-Instruct-v0.1:5S & \colorbox[RGB]{255, 204, 229}{Inval.} & 0.328 & 0.134 & \colorbox[RGB]{255, 204, 229}{Inval.} & 0.147 & 0.459 & 0.595 & \colorbox[RGB]{255, 204, 229}{Inval.} & 0.891 & \colorbox[RGB]{255, 204, 229}{Inval.}\\

& Gemma 2-9b-it:5S & \colorbox[RGB]{255, 204, 229}{Inval.} & \colorbox[RGB]{255, 204, 229}{Inval.} & \colorbox[RGB]{255, 204, 229}{Inval.} & \colorbox[RGB]{255, 204, 229}{Inval.} & 1.013 & 0.674 & 0.919 & \colorbox[RGB]{255, 204, 229}{Inval.} & 1.194 & 0.626\\
 
& MatBERT-109M & 7.651 & 6.083 & 15.558 & 1.558 & 9.976 & 46.586 & \bf 11.027 & \bf 3.055 & \colorbox[RGB]{204, 229, 255}{\textbf{2.152}} & \colorbox[RGB]{204, 229, 255}{\textbf{1.847}} \\

& LLM-Prop-35M & \colorbox[RGB]{204, 229, 255}{\textbf{9.116}} & \colorbox[RGB]{204, 229, 255}{\textbf{7.204}} & \colorbox[RGB]{204, 229, 255}{\textbf{16.224}} & \colorbox[RGB]{204, 229, 255}{\textbf{1.706}} & \colorbox[RGB]{204, 229, 255}{\textbf{15.926}} & \bf 93.001 & 9.995 & 3.016 & 1.950 & 1.656 \\ 

\bottomrule
 
\end{tabular}}
\label{tab:results_regression}
\end{table}

\begin{table}[t]
\caption{The Wtd. Avg. AUC scores (the higher the better) for the classification tasks in the LLM4Mat-Bench.}

\centering

\begin{tabular}{l l c c  }

\toprule

\multirow{3}{*}{\textbf{Input} }& \multirow{3}{*}{\textbf{Model}} &  \multicolumn{2}{c}{\bf Dataset}\\
\cmidrule(r){3-4}
 &  & \textbf{MP} & \textbf{SNUMAT} \\
& & 2 tasks & 3 tasks\\
 
 \midrule

 CIF & CGCNN (baseline) & 0.846 & 0.722  \\
 
 \midrule

 \multirow{6}{*}{Comp.} & Llama 2-7b-chat:0S &  0.491  & \colorbox[RGB]{255, 204, 229}{Inval.}  \\
 
 & Llama 2-7b-chat:5S & 0.507 & 0.466\\

 & Mistral 7b-Instruct-v0.1:5S & 0.506 & 0.488\\
 
 & Gemma 2-9b-it:5S & 0.513 & 0.584\\
 
 & MatBERT-109M & \textbf{0.722} & 0.712 \\
 
 & LLM-Prop-35M & 0.691 & \bf 0.716 \\
 
 \midrule

\multirow{6}{*}{CIF} & Llama 2-7b-chat:0S & 0.501 & 0.489 \\

& Llama 2-7b-chat:5S & 0.502 & 0.474  \\

& Mistral 7b-Instruct-v0.1:5S & \colorbox[RGB]{255, 204, 229}{Inval.} & \colorbox[RGB]{255, 204, 229}{Inval.}\\

& Gemma 2-9b-it:5S & \colorbox[RGB]{255, 204, 229}{Inval.} & 0.522\\
 
& MatBERT-109M & \colorbox[RGB]{204, 229, 255}{\bf 0.750} & \bf 0.717  \\

& LLM-Prop-35M & 0.738 & 0.660\\

\midrule

\multirow{6}{*}{Descr.} & Llama 2-7b-chat:0S & 0.500 & \colorbox[RGB]{255, 204, 229}{Inval.} \\

& Llama 2-7b-chat:5S & 0.502 & 0.568 \\

& Mistral 7b-Instruct-v0.1:5S & 0.499 & 0.501\\

& Gemma 2-9b-it:5S & \colorbox[RGB]{255, 204, 229}{Inval.} & 0.492\\
 
& MatBERT-109M & 0.735 & 0.730 \\

& LLM-Prop-35M & \bf 0.742 & \colorbox[RGB]{204, 229, 255}{\textbf{0.735}} \\ 

\bottomrule
 
\end{tabular}
\label{tab:results_classification}
\end{table}

\begin{figure*}[t]
    \centering
    \includegraphics[width=\linewidth]{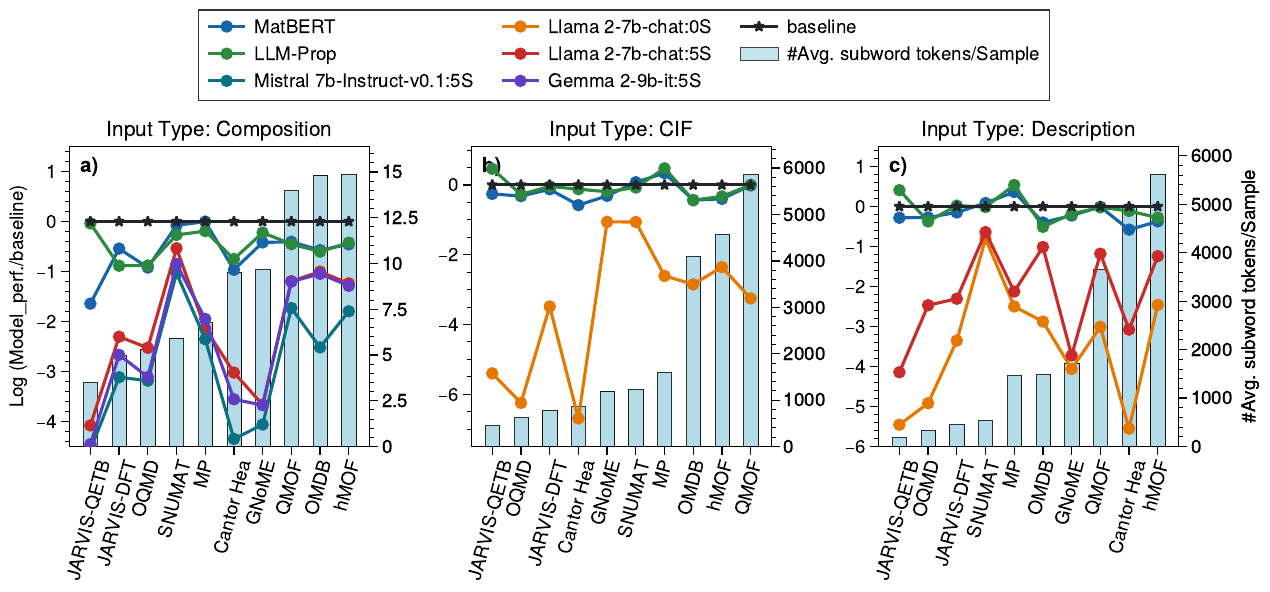}
    \caption{The performance comparison across models for each material representation is presented. The left y-axis shows the log-normalized performance of each LLM-based model relative to the baseline (CGCNN), while the right y-axis (bar plots) displays the average subword tokens per sample for each dataset. Datasets on the x-axis are ordered by increasing average subword tokens. Results for some chat-like models  are missing in each subplot due to invalid outputs on at least one of the property. Higher values in the line plots indicate better performance. Panels (a), (b), and (c) represents the performance comparison where the input is a chemical composition, CIF, and structure description, respectively.}
    \label{fig:model_comparison}
\end{figure*}

\begin{figure*}[t]
    \centering
    \includegraphics[width=\linewidth]{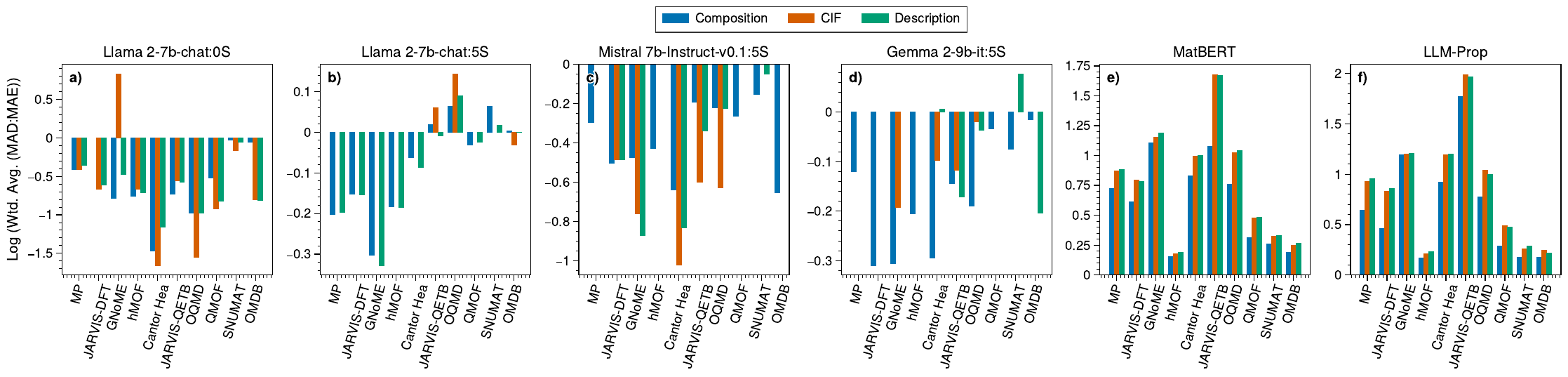}
    \caption{The performance comparison across material representations for each LLM-based model is shown. The y-axis represents the log-normalized Weighted Average (MAD:MAE) score for each representation, while the x-axis displays randomly ordered datasets. In the (a)-(d) plots, some Composition and Structure performance results are missing due to invalid outputs. A higher y-axis value indicates better performance. Panels (a) to (f) represents the results for Llama 2-7b-chat:0S, Llama 2-7b-chat:5S, Mistral 7b-Instruct-v0.1:5S, Gemma 2-9b-it:5S, MatBERT, and LLM-Prop, respectively.}
    \label{fig:input_comparison}
\end{figure*}

\subsection{Discussion}\label{sec:discussion}
Table \ref{tab:results_regression} and \ref{tab:results_classification}, and Figure \ref{fig:model_comparison} and \ref{fig:input_comparison} show the main results. The detailed results on each dataset can be found in Appendix \ref{app:results_per_each_dataset}. The main observations are as follows:

\textbf{Small, task-specific predictive LLMs exhibit significantly better performance than larger, generative general-purpose LLMs.} This performance disparity is evident across both regression (Table \ref{tab:results_regression} and Figure \ref{fig:model_comparison}) and classification tasks (Table \ref{tab:results_classification}) on all 10 datasets. Specifically, LLM-Prop and MatBERT outperform conversational LLMs by a substantial margin, despite being approximately 200 and 64 times smaller in size, respectively. In regression tasks, LLM-Prop achieves the highest accuracy on 8 out of 10 datasets, with MatBERT leading on the remaining 2 datasets. For classification tasks, both LLM-Prop and MatBERT deliver the best performance on 1 out of 2 datasets. LLM-Prop surpasses MatBERT by 1.8\% on the SNUMAT dataset, whereas MatBERT outperforms LLM-Prop by 0.8\% on the other dataset. As anticipated, a modest enhancement in average performance is observed across various datasets and input formats when the Llama 2-7b-chat model is evaluated using 5-shot prompts rather than 0-shot prompts. Determining the optimal number of examples required to achieve peak performance will be the focus of future work. 

\textbf{General-purpose generative  LLMs hallucinate and often fail to generate valid property values.} As shown in Table \ref{tab:results_regression}, Table \ref{tab:results_classification}, Figure \ref{fig:input_comparison}, and Appendix \ref{app:results_per_each_dataset}, Llama, Gemma, and Mistral models produce invalid outputs on multiple tasks, where the expected property value is missing. This issue occurs less frequently when the input is a description or chemical formula, but more commonly when the input is a CIF file. One reason may be that descriptions and chemical formulas resemble natural language, which LLMs can more easily interpret compared to CIF files. This may also indicate that when the input modality during inference differs significantly from the modalities encountered during pretraining, fine-tuning is necessary to achieve reasonable performance. Another key observation is that these models often generate the same property value for different inputs (i.e. hallucinate), contributing to their poor performance across multiple tasks. These findings highlight the importance of caution when using general-purpose generative LLMs for materials property prediction and emphasize the need for fine-tuned, task-specific LLM-based models.

\textbf{Representing materials with their textual descriptions improves the performance of LLM-based property predictors compared to other representations.} We observe a significant performance improvement when the input is a description compared to when it is a CIF file or a chemical formula. One of the possible reasons for this might be that LLMs are more adept at learning from natural language data. On the other hand, although material compositions appear more natural to LLMs compared to CIF files, they lack sufficient structural information. This is likely why LLMs with CIF files as input significantly outperform those using chemical formulas.

\textbf{More advanced, general-purpose generative LLMs do not necessarily yield better results in predicting material properties.} In Figure \ref{fig:chat-llms-comparison}, we compare the performance of Llama 2-7b-chat-hf model with advanced versions of Llama of comparable sizes when predicting material's band gap and its stability. Similar comparisons are also conducted for the Mistral   \citep{jiang2023mistral} and Gemma   \citep{team2024gemma} models. The results indicate that, despite being trained on substantially larger and higher-quality datasets, more advanced versions of generative LLMs show limited improvements in performance and validity of predictions for material properties. For instance, Llama 3 and 3.1 8b models were trained on over 15 trillion tokens—around eight times more data than the 2 trillion tokens used for the Llama 2 7b models. This finding highlights the ongoing challenges of leveraging LLMs in material property prediction and underscores the need for further research to harness the potential of these robust models in this domain.

\textbf{The performance on energetic properties is consistently better across all datasets compared to other properties.} This is consistent with the trend observed in the community benchmarks such as MatBench and JARVIS-Leaderboard, where energetic properties are among those that can be most accurately predicted~\citep{dunn2020benchmarking,choudhary2024jarvis}. This is not surprising because energy is known to be relatively well predicted from e.g., compositions and atom coordination (bonding), which is inherently represented in GNNs and also presented in text descriptions.

\textbf{Task-specific predictive LLM-based models excel with shorter textual descriptions, while CGCNN performs better on datasets with longer descriptions.} While the focus on this work is on LLMs, a comparison with a simple but widely used GNN-based baseline suggests room for improvement in LLM-based property prediction. For regression tasks, LLM-Prop outperforms CGCNN on only 4 out of 10 datasets (MP, JARVIS-DFT, JARVIS-QETB, and SNUMAT), and MatBERT outperforms CGCNN on just 2 out of 10 datasets (MP and JARVIS-QETB). In contrast, CGCNN achieves the best performance on 5 out of 10 datasets (GNoME, hMOF, Cantor HEA, OQMD, and OMDB). Further analysis reveals that CGCNN tends to perform better than LLM-based models on datasets with relatively longer textual descriptions, while LLM-based models excel on datasets with shorter descriptions (see Table \ref{llm4matbench_stats}). The performance gain on shorter descriptions may stem from LLM-based models’ ability to leverage more context from compact text, while CGCNN consistently benefits from training on the entire crystal structure.

\section{Conclusion} \label{sec:conclusion}

LLMs are increasingly being utilized in materials science, particularly for materials property prediction and discovery. However, the absence of standardized evaluation benchmarks has impeded progress in this field. We introduced LLM4Mat-Bench, a comprehensive benchmark dataset designed to evaluate LLMs for predicting properties of atomic and molecular crystals and MOFs. Our results demonstrate the limitations of general-purpose LLMs in this domain and underscore the necessity for task-specific predictive models and instruction-tuned LLMs tailored for materials property prediction. These findings emphasize the importance of using LLM4Mat-Bench to advance the development of more effective LLMs in materials science.

\section{Limitations} \label{sec:limitations}

Due to computational constraints and the number of experiments, we were unable to conduct thorough hyperparameter searches for each property and dataset. The reported settings were optimized on the MP dataset and then fixed for other datasets. For each model, we highlighted hyperparameter settings that may improve performance (see Section \ref{sec:models_info}). Additionally, we could not include results from SOTA commercial LLMs such as GPT-4o\footnote{https://openai.com/index/hello-gpt-4o/} or Claude 3.5 Sonnet\footnote{https://www.anthropic.com/news/claude-3-5-sonnet} due to budget constraints. 

We also encountered issues with chat-based models, which sometimes failed to follow the output format, producing invalid or incomplete outputs. Extracting property values was therefore challenging. We believe further instruction-tuning chat-based models on the provided prompts could mitigate these issues. 

Furthermore, we did not include comparisons with dataset-specific retrieval-augmented generation (RAG) models, such as the recently developed LLaMP   \citep{chiang2024llamp}, a RAG-based model tailored for interaction with the MP dataset. Our work aims to provide a comprehensive benchmark and baseline results to advance the evaluation of LLM-based methods for materials property prediction. Future work should address these limitations.

\subsection*{Acknowledgments}
Adji Bousso Dieng acknowledges support from the National Science Foundation, Office of Advanced Cyberinfrastructure (OAC) \#2118201, and from the Schmidt Sciences AI2050 Early Career Fellowship. 

\bibliographystyle{apa}
\bibliography{LLM4Mat-Bench-Arxiv/arxiv}

\clearpage

\section*{Appendices}
\appendix

\subsection*{Table of Contents}
\startcontents[sections]
\printcontents[sections]{l}{1}{\setcounter{tocdepth}{2}}

\clearpage
\section{Materials Representations}
\begin{table}[hbt!]
    \centering
    \caption{\label{tab:llm4matbench_input_formats} LLM4Mat-Bench material representations of Sodium Chloride (NaCl).}
    \begin{tabular}{p{12.0cm}}
        \toprule
         \textbf{Crystal Information File (CIF)} \\ 
        
        \begin{lstlisting}
# generated using pymatgen
data_NaCl
_symmetry_space_group_name_H-M   'P 1'
_cell_length_a   3.50219000
_cell_length_b   3.50219000
_cell_length_c   3.50219000
_cell_angle_alpha   90.00000000
_cell_angle_beta   90.00000000
_cell_angle_gamma   90.00000000
_symmetry_Int_Tables_number   1
_chemical_formula_structural   NaCl
_chemical_formula_sum   'Na1 Cl1'
_cell_volume   42.95553287
_cell_formula_units_Z   1
loop_
 _symmetry_equiv_pos_site_id
 _symmetry_equiv_pos_as_xyz
  1  'x, y, z'
loop_
 _atom_type_symbol
 _atom_type_oxidation_number
  Na+  1.0
  Cl-  -1.0
loop_
 _atom_site_type_symbol
 _atom_site_label
 _atom_site_symmetry_multiplicity
 _atom_site_fract_x
 _atom_site_fract_y
 _atom_site_fract_z
 _atom_site_occupancy
  Na+  Na0  1  0.00000000  0.00000000  0.00000000  1
  Cl-  Cl1  1  0.50000000  0.50000000  0.50000000  1
        \end{lstlisting}\\
        \midrule
        \textbf{Description} \\ \\
           NaCl is Tetraauricupride structured and crystallizes in the cubic $P\overline{m}3m$ space group. Na\textsuperscript{1+} is bonded in a body-centered cubic geometry to eight equivalent Cl\textsuperscript{1-} atoms. All Na-Cl bond lengths are 3.03 Å. Cl\textsuperscript{1-} is bonded in a body-centered cubic geometry to eight equivalent Na\textsuperscript{1+} atoms.\\
        
        \bottomrule
    \end{tabular}
\end{table}

\clearpage
\section{Statistics of All Properties in LLM4Mat-Bench}

\begin{table}[hbt!]
\centering
\caption{Statistics of all datasets in LLM4Mat-Bench. It is important to note that we retain the naming convention of each property from the original data source with the intent to provide the distribution of properties in each dataset.}

\resizebox{1.0\textwidth}{!}{
\begin{tabular}{l l c c c c c c c c c c}
\toprule \multirow{2}{*}{\bf Property}  & \multirow{2}{*}{\bf Task type} & \multicolumn{10}{c}{\bf \# Samples/Data source} \\ 

\cmidrule(r){3-12}

& & \bf JARVIS-DFT & \bf Materials Project & \bf SNUMAT & \bf hMOF & \bf GNoME & \bf JARVIS-QETB & \bf Cantor HEA & \bf QMOF & \bf OQMD & \bf OMDB \\ \midrule

Bandgap &  Regression &  - & 145,302 & - & - & 288,209 & - & - & 16,340 & 1,007,324 & 12,500\\
Bandgap (OPT) & Regression & 75,965 & - & - & - & - & - & - & - & - & -\\
Bandgap (MBJ) & Regression & 19,800 & - & - & - & - & - & - & - & - & -\\
Bandgap GGA & Regression & - & - & 10,481 & - & - & - & - & - & - & -\\
Bandgap HSE & Regression & - & - & 10,481 & - & - & - & - & - & - & -\\
Bandgap GGA Optical & Regression & - & - & 10,481 & - & - & - & - & - & - & -\\
Bandgap HSE Optical & Regression & - & - & 10,481 & - & - & - & - & - & - & -\\
Indirect Bandgap & Regression & - & - & - & - & - & 829,576 & - & - & - & -\\
Formation Energy Per Atom (FEPA) & Regression & 75,965 & 145,262 & - & - & 384,871 & 829,576 & 84,024 & - & 1,008,266 & - \\
Energy Per Atom (EPA) & Regression & - & 145,262 & - & - & - & 829,576 & 84,024 & - & - & -\\
Decomposition Energy Per Atom (DEPA) & Regression & - & - & - & - & 384,871 & - & - & - & - & - \\
Energy Above Hull (Ehull)  & Regression & 75,965 & 145,262 & - & - & - & - & 84,024 & - & - & -\\
Total Energy & Regression & 75,965 & - & - & - & 384,871 & 829,576 & - & 16,340 & - & - \\
Efermi & Regression & - & 145,262 & - & - & - & - & - & - & - & -\\
Exfoliation Energy & Regression & 812 & - & - & - & - & - & - & - & - & -\\
Bulk Modulus (Kv) & Regression & 23,823 & - & - & - & - & - & - & - & - & -\\
Shear Modulus (Gv) & Regression & 23,823 & - & - & - & - & - & - & - & - & -\\
SLME & Regression & 9,765 & - & - & - & - & - & - & - & - & -\\
Spillage & Regression & 11,377 & - & - & - & - & - & - & - & - & -\\
$\epsilon_{x}$ (OPT) & Regression & 52,158 & - & - & - & - & - & - & - & - & - \\
$\epsilon$ (DFPT) & Regression & 4,704 & - & - & - & - & - & - & - & - & - \\
Max. piezoelectri c strain coeff (dij) & Regression & 3,347 & - & - & - & - & - & - & - & - & -\\
Max. piezo. stress coeff (eij) & Regression & 4,797 & - & - & - & - & - & - & - & - & -\\
Max. EFG & Regression & 11,871 & - & - & - & - & - & - & - & - & -\\
Avg. $m_{e}$  & Regression & 17,643 & - & - & - & - & - & - & - & - & -\\
Is Stable & Classification & - & 145,262 & - & - & - & - & - & - & - & -\\
Is Gap Direct & Classification & - & 145,262 & - & - & - & - & - & - & - & -\\
n-Seedbeck & Regression & 23,211 & - & - & - & - & - & - & - & - & -\\
n-PF & Regression & 23,211 & - & - & - & - & - & - & - & - & - \\
p-Seedbeck & Regression & 23,211 & - & - & - & - & - & - & - & - & -\\
p-PF & Regression & 23,211 & - & - & - & - & - & - & - & - & -\\ 
Density & Regression & - & 145,262 & - & - & 384,871 & - & - & - & - & - \\
Density Atomic & Regression & - & 145,262 & - & - & - & - & - & - & - & -\\
Volume & Regression & - & 145,262 & - & - & 384,871 & - & - & - & - & - \\
Volume Per Atom (VPA) & Regression & - & - & - & - & - & - & 84,024 & - & - & -\\
Is Direct & Classification  & - & - & 10,481 & - & - & - & - & - & - & -\\
Is Direct HSE & Classification & - & - & 10,481 & - & - & - & - & - & - & -\\
SOC & Classification & - & - & 10,481 & - & - & - & - & - & - & -\\
LCD & Regression & - & - & - & 133,524 & - & - & - & 16,340 & - & -\\
PLD & Regression & - & - & - & 133,524 & - & - & - & 16,340 & - & -\\
Max CO2 & Regression & - & - & - & 133,524 & - & - & - & - & - & -\\
Min CO2 & Regression & - & - & - & 133,524 & - & - & - & - & - & -\\
Void Fraction & Regression & - & - & - & 133,524 & - & - & - & - & - & -\\
Surface Area m2g & Regression & - & - & - & 133,524 & - & - & - & - & - & -\\
Surface Area m2cm3 & Regression & - & - & - & 133,524 & - & - & - & - & - & -\\

\bottomrule
\end{tabular}}

\label{llm4matbench_prop_stats}
\end{table}

\section{Chat-like Model Inference Details}
\label{sec:inference_details}

\begin{table}[!htbp]
    \centering
    \caption{Hyperparameters used during inference. Temp. represents temperature.}
    \label{tab:inference_details}
    \resizebox{1.0\textwidth}{!}{
    \begin{tabular}{l l l c c c c}
        \toprule
        \bf Model Type & \bf Model Name & \bf Input Length & \bf Output Length & \bf Batch Size & \bf Temp. & \bf Top\_K \\
        \midrule
         \multirow{4}{*}{\makecell{Llama \\ Family}} & Llama 2-7b-chat-hf & 4000 & 256 & 256 & 0.8 & 10 \\
         & Llama 3-8b-Instruct & 8000 & 256 & 256 & 0.8 & 10  \\
         & Llama 3.1-8b-Instruct & 98000 & 256 & 128 & 0.8 & 10  \\
         & Llama 3.2-3b-Instruct & 47000 & 256 & 128 & 0.8 & 10  \\
         \midrule
         \multirow{3}{*}{\makecell{Gemma \\ Family}} & Gemma 7b-it & 4000 & 256 & 256 & 0.8 & 10  \\
         & Gemma 1.1-7b-it & 4000 & 256 & 256 & 0.8 & 10  \\
         & Gemma 2-9b-it & 3000 & 256 & 256 & 0.8 & 10  \\
         \midrule
         \multirow{3}{*}{\makecell{Mistral \\ Family}} & Mistral 7b-Instruct-v0.1 & 20000 & 256 & 256 & 0.8 & 10  \\
         & Mistral 7b-Instruct-v0.2 & 20000 & 256 & 256 & 0.8 & 10  \\
         & Mistral 7b-Instruct-v0.3 & 20000 & 256 & 256 & 0.8 & 10  \\
         \bottomrule
    \end{tabular}
    }
\end{table}

\clearpage
\section{Prompt Templates}
\label{llama_prompt_templates}

\begin{figure}[!htbp]
    \centering
    \includegraphics[width=\linewidth]{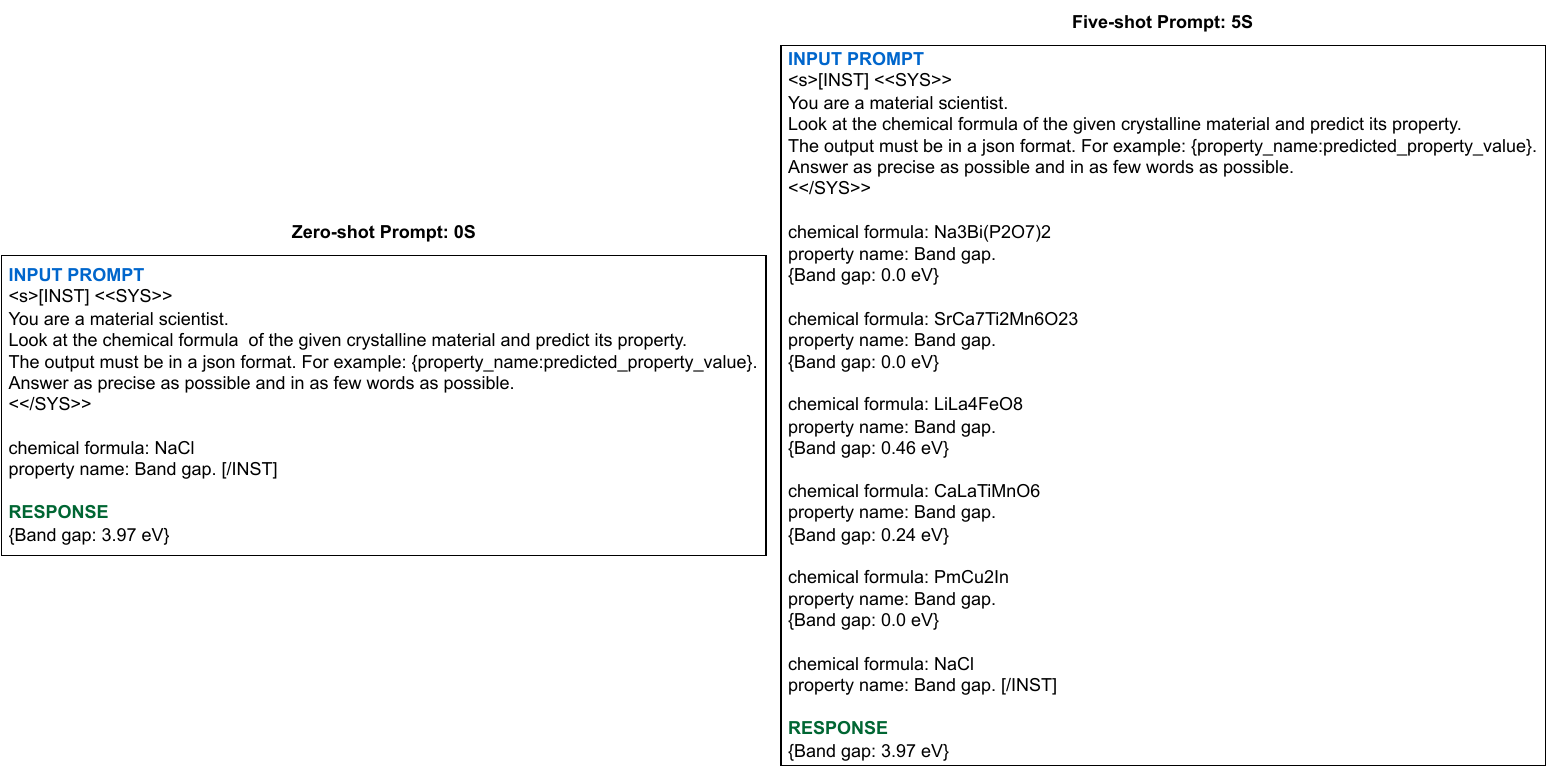}
    \caption{Prompt templates when the input is a chemical formula.}
    \label{fig:prompt_template}
\end{figure}
\FloatBarrier

\begin{figure}[!htbp]
    \centering
    \includegraphics[width=\linewidth]{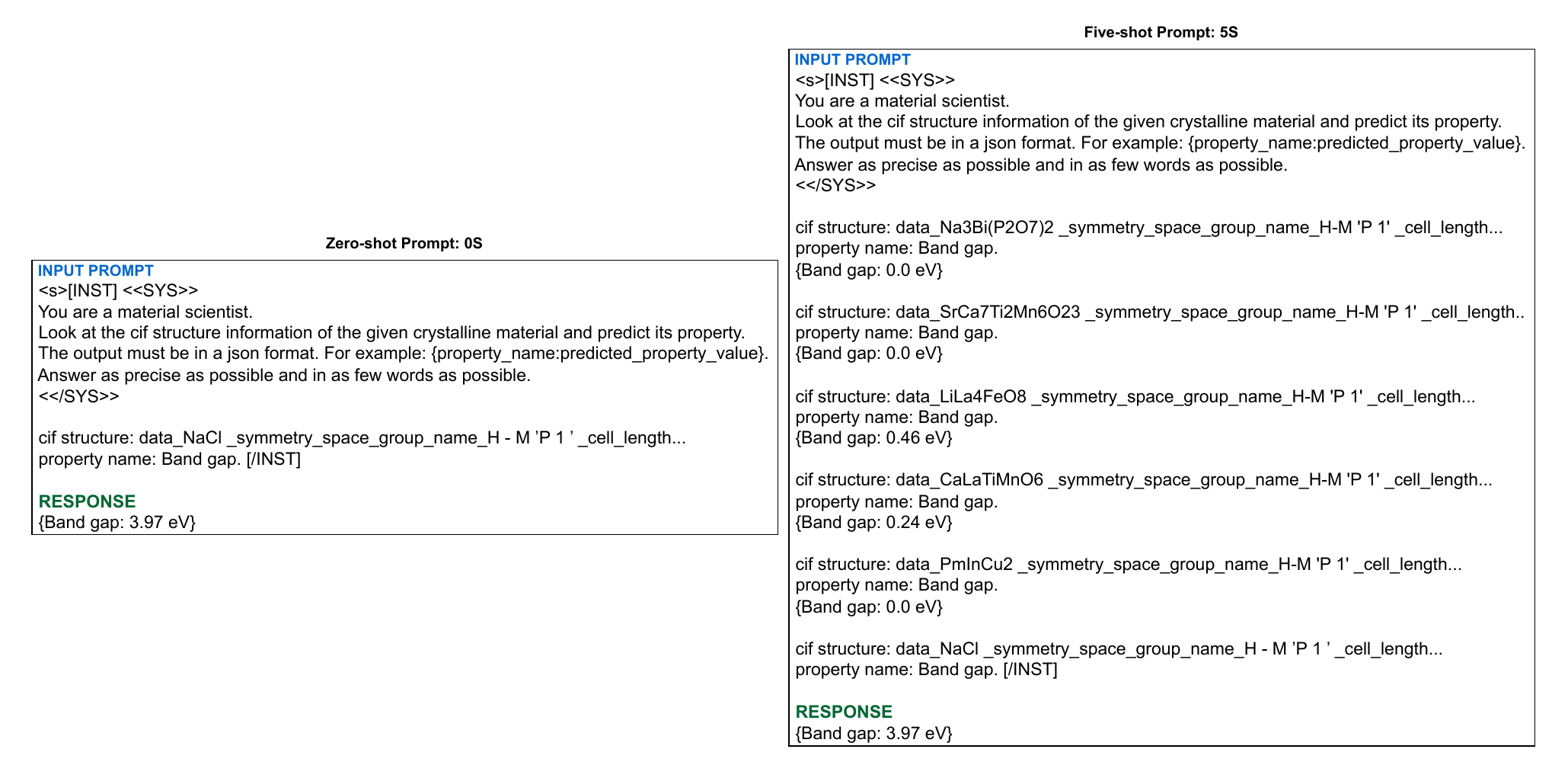}
    \caption{Prompt templates when the input is a CIF file.}
    \label{fig:prompt_template}
\end{figure}
\FloatBarrier

\begin{figure}[!htbp]
    \centering
    \includegraphics[width=\linewidth]{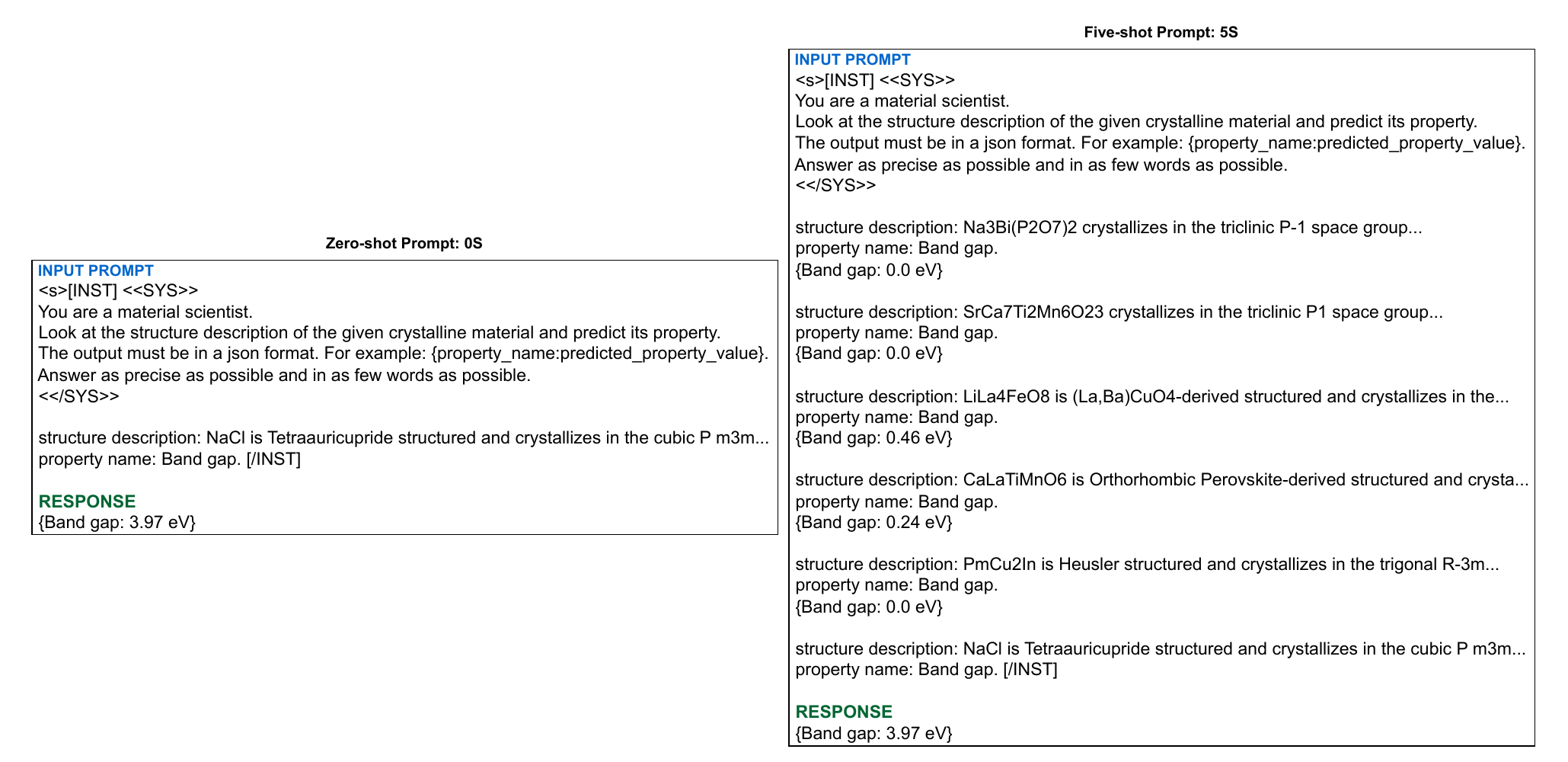}
    \caption{Prompt templates when the input is a crystal structure description.}
    \label{fig:prompt_template}
\end{figure}

\clearpage
\section{Result Details for Each Dataset}
\label{app:results_per_each_dataset}

\subsection{Detailed MAD:MAE Results}

\begin{table*}[hbt!]
\centering
\caption{\label{tab:mp_results} Results for MP dataset. The performance on regression tasks is evaluated in terms of MAD:MAE ratio (the higher the better) while that of classification tasks (Is Stable and Is Gab Direct) is evaluated in terms of AUC score. FEPA: Formation Energy Per Atom, EPA: Energy Per Atom.}
\resizebox{1.0\textwidth}{!}{
\begin{tabular}{l l c c c c c c c c c c}

\toprule

\multirow{2}{*}{\textbf{Input}} & \multirow{2}{*}{\textbf{Model}} & \textbf{FEPA} & \textbf{Bandgap} & \textbf{EPA} & \textbf{Ehull} & \textbf{Efermi} & \textbf{Density} & \textbf{Density Atomic} & \textbf{Volume} & \textbf{Is Stable} & \textbf{Is Gab Direct} \\
& & \bf 145.2K & \bf 145.3K & \bf 145.2K & \bf 145.2K & \bf 145.2K & \bf 145.2K & \bf 145.2K & \bf 145.2K & \bf 145.2K & \bf 145.2K \\
 
 \midrule

 CIF & CGCNN (baseline) & 8.151 & 3.255 & 7.224 & 3.874 & 3.689 & 8.773 & 5.888 & 1.703 & 0.882 & 0.810 \\
 
 \midrule

 \multirow{6}{*}{Comp.} & Llama 2-7b-chat:0S &  0.008 & 0.623 & 0.009 & 0.001 & 0.003 & 0.967 & 0.754 & 0.747 & 0.500 & 0.482 \\
 
 & Llama 2-7b-chat:5S & 0.33 & 1.217 & 0.239 & 0.132 & 0.706 & 0.899 & 0.724 & 0.771 & 0.502 & 0.512\\

 & Mistral 7b-Instruct-v0.1:5S & 0.109 & 0.898 & 0.345 & 0.164 & 0.282 & 0.255 & 1.01 & 0.973 & 0.505 & 0.508\\

 & Gemma 2-9b-it:5S & 0.519 & 0.999 & 0.482 & 0.088 & 0.669 & 1.368 & 0.818 & 1.118 & 0.512 & 0.513\\

 & MatBERT-109M & \bf 8.151 & \bf 2.971 & \bf 9.32 & \bf 2.583 & \bf 3.527 & \bf 7.626 & \bf 5.26 & \bf 3.099 & \textbf{0.764} & \textbf{0.681} \\
\bottomrule
 
\end{tabular}}
\end{table*}

\begin{table*}[hbt!]
\centering
\caption{\label{tab:jarvis_dft_results} Results for JARVIS-DFT. he performance on regression tasks is evaluated in terms of MAD:MAE ratio (the higher the better). FEPA: Formation Energy Per Atom, Tot. En.: Total Energy, Exf. En.: Exfoliation Energy.}
\resizebox{1.0\textwidth}{!}{
\begin{tabular}{l l c  c  c  c  c  c  c  c  c  c }

\toprule

\multirow{2}{*}{\textbf{Input}} & \multirow{2}{*}{\textbf{Model}} & \textbf{FEPA  } & \textbf{Bandgap (OPT)  } & \textbf{Tot. En.  } & \textbf{Ehull  } & \textbf{Bandgap (MBJ)  } & \textbf{Kv  } & \textbf{Gv  } & \textbf{SLME  } & \textbf{Spillage  } & \textbf{$\epsilon_{x}$ (OPT)  }\\

& & \bf 75.9K & \bf 75.9K & \bf 75.9K & \bf 75.9K & \bf 19.8K & \bf 23.8K & \bf 23.8K & \bf 9.7K & \bf 11.3K & \bf 18.2K \\
 \midrule
 
 CIF & CGCNN (baseline) & 13.615 & 4.797 & 22.906 & 1.573 & 4.497 & 3.715 & 2.337 & 1.862 & 1.271 & 2.425 \\
 
 \midrule

\multirow{6}{*}{Comp.} & Llama 2-7b-chat:0S & 0.021 & 0.011 & 0.02 & 0.005 & 0.92 & 0.428 & 0.374 & 0.148 & \colorbox[RGB]{255, 204, 229}{Inval.} & 0.18  \\

& Llama 2-7b-chat:5S & 0.886 & 0.011 & 0.02 & 1.292 & 0.979 & 0.88 & 0.992 & 0.456 & 0.85 & 1.148 \\

& Mistral 7b-Instruct-v0.1:5S & 0.128 & 0.011 & 0.019 & 0.171 & 0.513 & 0.878 & 0.93 & 0.369 & 0.123 & 0.996\\

& Gemma 2-9b-it:5S & 0.15 & 0.011 & 0.02 & 0.601 & 0.973 & 1.257 & 1.137 & 0.691 & 0.261 & 1.174\\

& MatBERT-109M &  \bf 6.808 & \bf 4.083 & \bf 9.21 & \bf 2.786 & \bf 3.755 & \bf 2.906 & \bf 1.928 & \bf 1.801 & \bf 1.243 & \bf 2.017 \\
\bottomrule
 
\end{tabular}
}

\end{table*}

\begin{table*}[hbt!]
\centering
\caption{\label{tab:snumat_results} Results for SNUMAT. The performance on regression tasks is evaluated in terms of MAD:MAE ratio (the higher the better) while that of classification tasks (Is Direct, Is Direct HSE, and SOC) is evaluated in terms of AUC score.}
\resizebox{1.0\textwidth}{!}{
\begin{tabular}{l l c  c  c  c  c  c  c }

\toprule

\multirow{2}{*}{\textbf{Input}} & \multirow{2}{*}{\textbf{Model}} & \textbf{Bandgap GGA } & \textbf{Bandgap HSE } & \textbf{Bandgap GGA Optical } & \textbf{Bandgap HSE Optical } & \textbf{Is Direct } & \textbf{Is Direct HSE } & \textbf{SOC}\\

& & \bf 10.3K & \bf 10.3K & \bf 10.3K & \bf 10.3K & \bf 10.3K & \bf 10.3K & \bf 10.3K \\

 \midrule
 
 CIF & CGCNN (baseline) & 2.075 & 2.257 & 1.727 & 1.835 & 0.691 & 0.675 & 0.800 \\
 
 \midrule

\multirow{6}{*}{Comp.} & Llama 2-7b-chat:0S & 0.797 & 0.948 & 1.156 & 0.859  & 0.503 & 0.484 & \colorbox[RGB]{255, 204, 229}{Inval.}\\

& Llama 2-7b-chat:5S & 1.267 & 1.327 & 0.862 & 1.174 & 0.475 & 0.468 & 0.455 \\

& Mistral 7b-Instruct-v0.1:5S &  0.794 & 0.468 & 1.011 & 0.539 & 0.466 & 0.499 & 0.5\\

& Gemma 2-9b-it:5S & 0.741 & 1.168 & 0.615 & 0.848 & 0.530 &  0.485 & 0.719\\
 
& MatBERT-109M & \bf 1.899 & \bf 1.975 & \bf 1.646 & \bf 1.793  & \textbf{0.671} & \textbf{0.645} & 0.820 \\
\bottomrule
 
\end{tabular}}
\end{table*}

\begin{table*}[hbt!]
\centering
\caption{\label{tab:gnome_results} Results for GNoME. The performance on regression tasks is evaluated in terms of MAD:MAE ratio (the higher the better). FEPA: Formation Energy Per Atom, DEPA: Decomposition Energy Per Atom, Tot. En.: Total Energy.}
\resizebox{1.0\textwidth}{!}{
\begin{tabular}{l l c  c  c  c  c  c }

\toprule

\multirow{2}{*}{\textbf{Input}} & \multirow{2}{*}{\textbf{Model}} & \textbf{FEPA } & \textbf{Bandgap } & \textbf{DEPA } & \textbf{Tot. En. } & \textbf{Volume } & \textbf{Density } \\

& & \bf 376.2K & \bf 282.7K & \bf 376.2K & \bf 282.7K & \bf 282.7K & \bf 282.7K \\

 \midrule
 
 CIF & CGCNN (baseline) & 34.57 & 8.549 & 2.787 & 7.443 & 7.967 & 56.077 \\
 
 \midrule

\multirow{6}{*}{Comp.} & Llama 2-7b-chat:0S & 0.002 & 0.177 & 0.0 & 0.088 & 0.455 & 0.368  \\

& Llama 2-7b-chat:5S & 0.194 & 0.086 & 0.255 & 0.765 & 1.006 & 0.865 \\

& MatBERT-109M & \colorbox[RGB]{204, 229, 255}{\bf 30.248} & \bf 4.692 & \bf 2.787 & 8.57 & 13.157 & 15.145  \\

& Mistral 7b-Instruct-v0.1:5S & 0.192 & 0.242 & 0.022 & 0.664 & 0.858 & 0.186\\

& Gemma 2-9b-it:5S & 0.182 & 0.186 & 0.046 & 0.706 & 1.032 & 1.07\\
\bottomrule
 
\end{tabular}
}
\end{table*}

\begin{table*}[hbt!]
\centering
\caption{\label{tab:hmof_results} Results for hMOF. The performance on regression tasks is evaluated in terms of MAD:MAE ratio (the higher the better).}
\resizebox{1.0\textwidth}{!}{
\begin{tabular}{l l c  c  c  c  c  c  c}

\toprule

\multirow{2}{*}{\textbf{Input}} & \multirow{2}{*}{\textbf{Model}} & \textbf{Max CO2 } & \textbf{Min CO2 } & \textbf{LCD } & \textbf{PLD } & \textbf{Void Fraction } & \textbf{Surface Area m2g }  & \textbf{Surface Area m2cm3 }\\

& & \bf 132.7K & \bf 132.7K & \bf 132.7K & \bf 132.7K & \bf 132.7K & \bf 132.7K & \bf 132.7K \\

\midrule

 CIF & CGCNN (baseline) & 1.719 & 1.617 & 1.989 & 1.757 & 2.912 & 3.765 & 2.039 \\
 
\midrule

 \multirow{6}{*}{Comp.} & Llama 2-7b-chat:0S & 0.011 & 0.002 & 0.009 & 0.008 & 0.5 & 0.454 & 0.233 \\
 
& Llama 2-7b-chat:5S & 0.679 & 0.058 & 0.949 & 1.026 & 0.945 & 0.567 & 0.366 \\

& Mistral 7b-Instruct-v0.1:5S & 0.451 & 0.01 & 0.458 & 0.735 & 0.033 & 0.567 & 0.368\\

& Gemma 2-9b-it:5S & 0.679 & 0.049 & 0.971 & 0.973 & 0.765 & 0.567 & 0.364\\
 
& MatBERT-109M & 1.335 & \bf 1.41 & \bf 1.435 & 1.378 & 1.57 & 1.517 & \bf 1.367 \\
\bottomrule
 
\end{tabular}
}
\end{table*}

\begin{table}[t]
\centering
\caption{\label{tab:cantor_hea_results} Results for Cantor HEA. The performance on regression tasks is evaluated in terms of MAD:MAE ratio (the higher the better). FEPA: Formation Energy Per Atom, EPA:Energy Per Atom, VPA:Volume Per Atom.}

\begin{tabular}{l l c  c  c  c}

\toprule

\multirow{2}{*}{\textbf{Input}} & \multirow{2}{*}{\textbf{Model}} & \textbf{FEPA } & \textbf{EPA } & \textbf{Ehull } & \textbf{VPA }  \\

& & \bf 84.0K & \bf 84.0K & \bf 84.0K & \bf 84.0K \\

\midrule

 CIF & CGCNN (baseline) & 9.036 & 49.521 & 9.697 & 2.869 \\
 
 \midrule

\multirow{6}{*}{Comp.} & Llama 2-7b-chat:0S & 0.005 & 0.098 & 0.003 & 0.031 \\

& Llama 2-7b-chat:5S & 0.896 & 0.658 & 0.928 & 0.986 \\

& Mistral 7b-Instruct-v0.1:5S & 0.387 & 0.186 & 0.22 & 0.134\\

& Gemma 2-9b-it:5S & 0.148 & 0.65 & 0.964 & 0.268\\
 
& MatBERT-109M & \bf 3.286 & 16.17 & \bf 5.134 & 2.489 \\

& LLM-Prop-35M & \bf 3.286 & \bf 22.638 & \bf 5.134 & \bf 2.543  \\

\midrule

\multirow{6}{*}{CIF} & Llama 2-7b-chat:0S & 0.001 & 0.084 & 0.0 & 0.004 \\

& Llama 2-7b-chat:5S & \colorbox[RGB]{255, 204, 229}{Inval.} &  \colorbox[RGB]{255, 204, 229}{Inval.} & \colorbox[RGB]{255, 204, 229}{Inval.} & \colorbox[RGB]{255, 204, 229}{Inval.} \\

& Mistral 7b-Instruct-v0.1:5S & 0.012 & 0.21 & 0.144 & 0.015\\

& Gemma 2-9b-it:5S & 0.968 & 0.923 & 0.882 & 0.427 \\
 
& MatBERT-109M & 7.229 & 17.607 & 9.187 & 5.809  \\

& LLM-Prop-35M & \colorbox[RGB]{204, 229, 255}{\bf 8.341} & \colorbox[RGB]{204, 229, 255}{\bf 36.015} & \colorbox[RGB]{204, 229, 255}{\bf 11.636} & \bf 6.919 \\

\midrule

\multirow{6}{*}{Descr.} & Llama 2-7b-chat:0S & 0.001 & 0.101 & 0.164 & 0.011
 \\

& Llama 2-7b-chat:5S &0.797 & 0.615 & 0.938 & 0.93 \\

& Mistral 7b-Instruct-v0.1:5S & 0.125 & 0.068 & 0.377 & 0.018\\

& Gemma 2-9b-it:5S & 1.013 & \colorbox[RGB]{255, 204, 229}{Inval.} & \colorbox[RGB]{255, 204, 229}{Inval.} & \colorbox[RGB]{255, 204, 229}{Inval.}\\
 
& MatBERT-109M & 7.229 & 17.607 & 9.187 & 5.881 \\

& LLM-Prop-35M & \colorbox[RGB]{204, 229, 255}{\bf 8.341} & \colorbox[RGB]{204, 229, 255}{\bf 36.015} & \colorbox[RGB]{204, 229, 255}{\bf 11.636} & \colorbox[RGB]{204, 229, 255}{\bf 7.713} \\

\bottomrule
 
\end{tabular}
\end{table}

\begin{table}[t]
\centering
\caption{\label{tab:qmof_results} Results for QMOF. The performance on regression tasks is evaluated in terms of MAD:MAE ratio (the higher the better). Tot. En.: Total Energy.}

\begin{tabular}{l l c   c   c   c  }

\toprule

\multirow{2}{*}{\textbf{Input}} & \multirow{2}{*}{\textbf{Model}} & \textbf{Bandgap } & \textbf{Tot. En.} & \textbf{LCD } & \textbf{PLD }  \\

& & \bf 7.6K & \bf 7.6K & \bf 7.6K & \bf 7.6K \\

 \midrule

 CIF & CGCNN (baseline) & 2.431 & 1.489 & 4.068 & 4.317\\
 
 \midrule

\multirow{6}{*}{Comp.} & Llama 2-7b-chat:0S & 0.901 & 0.26 & 0.045 & 0.009
 \\

& Llama 2-7b-chat:5S & 0.648 & 0.754 & 1.241 & 1.086 \\

& Mistral 7b-Instruct-v0.1:5S & 0.236 & 0.885 & 0.807 & 0.254\\

& Gemma 2-9b-it:5S & 1.053 & 0.686 & 0.913 & 1.048\\
 
& MatBERT-109M & \bf 1.823 & \bf 1.695 & \bf 2.329 & \bf 2.349  \\
\bottomrule
 
\end{tabular}

\end{table}

\begin{table}[t]
\centering
\caption{\label{tab:jarvis_qetb_results} Results for JARVIS-QETB. The performance on regression tasks is evaluated in terms of MAD:MAE ratio (the higher the better). FEPA: Formation Energy Per Atom, EPA:Energy Per Atom, Tot. En.: Total Energy, Ind. Bandgap: Indirect Bandgap.}

\begin{tabular}{l l c  c  c  c  }

\toprule

\multirow{2}{*}{\textbf{Input}} & \multirow{2}{*}{\textbf{Model}} & \textbf{FEPA } & \textbf{EPA } & \textbf{Tot. En. } & \textbf{Ind. Bandgap }\\

& & \bf 623.9K & \bf 623.9K & \bf 623.9K & \bf 623.9K \\

 \midrule

 CIF & CGCNN (baseline) & 1.964 & 228.201 & 11.218 & 5.534 \\
 
 \midrule

 \multirow{6}{*}{Comp.} & Llama 2-7b-chat:0S & 0.003 & 0.369 & 0.172 & 0.21 \\
 
& Llama 2-7b-chat:5S & 0.812 & 1.037 & 1.032 & 1.306 \\

& Mistral 7b-Instruct-v0.1:5S & 0.693 & 0.878 & 0.608 & 0.394\\

& Gemma 2-9b-it:5S & 0.608 & 0.989 & 0.88 & 0.397\\
 
& MatBERT-109M & 1.431 & 37.979 & 8.19 & 0.21  \\

& LLM-Prop-35M & \bf 2.846 & \bf 211.757 & \bf 21.309 & \bf 1.861 \\

\midrule

\multirow{6}{*}{CIF} & Llama 2-7b-chat:0S & 0.003 & 0.412 & 0.656 & 0.04 \\ 

& Llama 2-7b-chat:5S & 0.8 & 1.024 & 1.076 & 1.71 \\

& Mistral 7b-Instruct-v0.1:5S & 0.085 & 0.507 & 0.212 & 0.199\\ 

& Gemma 2-9b-it:5S & 0.943 & 0.975 & 0.868 & 0.268\\
 
& MatBERT-109M & 24.72 & 135.156 & 26.094 & \bf 4.779 \\

& LLM-Prop-35M & 23.346 & \colorbox[RGB]{204, 229, 255}{\bf 318.291} & \colorbox[RGB]{204, 229, 255}{\bf 48.192} & 1.845 \\

\midrule

\multirow{6}{*}{Descr.} & Llama 2-7b-chat:0S & 0.003 & 0.408 & 0.484 & 0.16
 \\

& Llama 2-7b-chat:5S & 0.85 & 1.015 & 1.035 & 1.021 \\

& Mistral 7b-Instruct-v0.1:5S & 0.653 & 0.541 & 0.313 & 0.329\\

& Gemma 2-9b-it:5S & 0.659 & 0.891 & 0.876 & 0.27\\
 
& MatBERT-109M & \colorbox[RGB]{204, 229, 255}{\bf 26.265} & 122.884 & 29.409 & \colorbox[RGB]{204, 229, 255}{\bf 7.788}
 \\

& LLM-Prop-35M & 22.513 & 312.218 & 35.43 & 1.845 \\

\bottomrule
 
\end{tabular}
\end{table}

\begin{table}[t]
\centering
\caption{\label{tab:oqmd_results} Results for OQMD. The performance on regression tasks is evaluated in terms of MAD:MAE ratio (the higher the better). FEPA: Formation Energy Per Atom.}

\begin{tabular}{l l c  c }

\toprule

\multirow{2}{*}{\textbf{Input}} & \multirow{2}{*}{\textbf{Model}} & \textbf{FEPA } & \textbf{Bandgap }\\

& & \bf 963.5K & \bf 963.5K \\

 \midrule

 CIF & CGCNN (baseline) & 22.291 & 6.701 \\
 
 \midrule

  \multirow{6}{*}{Comp.} & Llama 2-7b-chat:0S & 0.019 & 0.192 \\

& Llama 2-7b-chat:5S & 1.013 & 1.306 \\

& Mistral 7b-Instruct-v0.1:5S & 0.723 & 0.479 \\

& Gemma 2-9b-it:5S & 0.682 & 0.613\\

& MatBERT-109M & 7.662 & \bf 3.883 \\

& LLM-Prop-35M & \bf 9.195 & 2.845 \\ 

\midrule

\multirow{6}{*}{CIF} & Llama 2-7b-chat:0S &  0.009 & 0.047\\

& Llama 2-7b-chat:5S & 1.051 & 1.731 \\

& Gemma 2-9b-it:5S & 0.248 & 0.224 \\
 
& Mistral 7b-Instruct-v0.1:5S & 0.815 & 1.099 \\

& MatBERT-109M & 13.879 & \colorbox[RGB]{204, 229, 255}{\bf 7.163} \\

& LLM-Prop-35M & \colorbox[RGB]{204, 229, 255}{\textbf{18.861}} & 3.22 \\

\midrule

\multirow{4}{*}{Descr.} & Llama 2-7b-chat:0S & 0.025 & 0.187 \\

& Llama 2-7b-chat:5S & 0.991 & 1.468 \\

& Mistral 7b-Instruct-v0.1:5S & 0.59 & 0.599 \\

& Gemma 2-9b-it:5S & 0.896 & 0.942\\

& MatBERT-109M & 15.012 & \bf 7.041 \\

& LLM-Prop-35M & \textbf{16.346} & 3.644 \\

\bottomrule
 
\end{tabular}
\end{table}

\begin{table}[t]
\centering
\caption{\label{tab:omdb_results} Results for OMDB. The performance on regression tasks is evaluated in terms of MAD:MAE ratio (the
higher the better).}
\begin{tabular}{l l c }

\toprule

\multirow{2}{*}{\textbf{Input}} & \multirow{2}{*}{\textbf{Model}} & \textbf{Bandgap }\\

& & \bf 12.1K \\

 \midrule

 CIF & CGCNN (baseline) & 2.751\\
 
 \midrule

 \multirow{6}{*}{Comp.} & Llama 2-7b-chat:0S & 0.886\\
& Llama 2-7b-chat:5S & 1.009\\
& Mistral 7b-Instruct-v0.1:5S & 0.222\\
& Gemma 2-9b-it:5S & 0.967\\
& MatBERT-109M & \textbf{1.554}\\
& LLM-Prop-35M & 1.507 \\

\midrule

\multirow{6}{*}{CIF} & Llama 2-7b-chat:0S & 0.159\\
& Llama 2-7b-chat:5S & 0.930\\
& Mistral 7b-Instruct-v0.1:5S & \colorbox[RGB]{255, 204, 229}{Inval.}\\
& Gemma 2-9b-it:5S & \colorbox[RGB]{255, 204, 229}{Inval.}\\
& MatBERT-109M & \textbf{1.777} \\
& LLM-Prop-35M  & \textbf{1.777} \\

\midrule

\multirow{6}{*}{Descr.} & Llama 2-7b-chat:0S & 0.155\\
& Llama 2-7b-chat:5S & 1.002 \\
& Mistral 7b-Instruct-v0.1:5S & \colorbox[RGB]{255, 204, 229}{Inval.}\\
& Gemma 2-9b-it:5S & 0.626\\
& MatBERT-109M & \colorbox[RGB]{204, 229, 255}{\textbf{1.847}}\\
& LLM-Prop-35M &  1.656 \\

\bottomrule

\end{tabular}
\end{table}

\clearpage
\subsection{Detailed MAE Results}

\begin{table}[hbt!]
\centering
\caption{Results for Materials Project (MP). The $\downarrow$ and $\uparrow$ on the right of each property denote the MAE error and AUC score, respectively. The Dummy model represents the MAD score that is calculated on the ground truth of each property.}

\resizebox{1.0\textwidth}{!}{
\begin{tabular}{l l c c c c c c c c c c | c c }

\toprule

\textbf{Input} & \textbf{Model} & \textbf{FEPA $\downarrow$} & \textbf{Bandgap $\downarrow$} & \textbf{EPA $\downarrow$} & \textbf{Ehull $\downarrow$} & \textbf{Efermi $\downarrow$} & \textbf{Density $\downarrow$} & \textbf{Density atomic $\downarrow$} & \textbf{Volume $\downarrow$} & \textbf{Is stable $\uparrow$} & \textbf{Is gab direct $\uparrow$}  & \textbf{Wtd. Avg.} & \textbf{Wtd. Avg.} \\
& & & & & & & & & & & & \bf (MAD:MAE) $\uparrow$ & \bf ROC AUC $\uparrow$\\
 
 \midrule

 - & Dummy (MAD) & 1.003 & 1.191 & 1.51 & 0.225 & 2.239 & 2.158 & 11.693 & 342.606 & - & - & - & - \\

 \midrule

 CIF & CGCNN (baseline) & 0.123 & 0.366 & 0.209 & 0.058 & 0.607 & 0.246 & 1.986 & 201.186 & 0.882 & 0.810 & 5.319 & 0.846\\
 
 \midrule

 \multirow{6}{*}{Comp.} & Llama 2-7b-chat:0S &  128.007  & 1.913  & 163.899  & 195.908  & 673.939 & 2.231 & 15.503 & 458.456 & 0.500 & 0.482 & 0.389 & 0.491 \\
 
 & Llama 2-7b-chat:5S & 3.034 & 0.979 & 6.330 & 1.706 & 3.172 & 2.400 & 16.150 & 444.378 & 0.502 & 0.512 & 0.627 & 0.507\\

& Mistral 7b-Instruct-v0.1:5S & 9.181 & 1.326 & 4.38 & 1.374 & 7.938 & 8.469 & 11.575 & 352.033 & 0.505 & 0.508 & 0.505 & 0.506\\
 
& Gemma 2-9b-it:5S & 1.932 & 1.192 & 3.135 & 2.567 & 3.347 & 1.578 & 14.298 & 306.371 & 0.512 & 0.513 & 0.758 & 0.513\\
\bottomrule
 
\end{tabular}}
\label{tab:mp_results}
\end{table}

\begin{table}[hbt!]
\centering
\caption{Results for JARVIS-DFT. The $\downarrow$ on the right of each property denotes the MAE error.}
\resizebox{1.0\textwidth}{!}{
\begin{tabular}{l l c  c  c  c  c  c  c  c  c  c  | c }

\toprule

\textbf{Input} & \textbf{Model} & \textbf{FEPA  $\downarrow$} & \textbf{Bandgap (OPT)  $\downarrow$} & \textbf{Total energy  $\downarrow$} & \textbf{Ehull  $\downarrow$} & \textbf{Bandgap (MBJ)  $\downarrow$} & \textbf{Kv  $\downarrow$} & \textbf{Gv  $\downarrow$} & \textbf{SLME  $\downarrow$} & \textbf{Spillage  $\downarrow$} & \textbf{$\epsilon_{x}$ (OPT)  $\downarrow$} & \textbf{Wtd. Avg.}  $\uparrow$\\
& & & & & & & & & & & & \bf (MAD:MAE) $\uparrow$ \\

 \midrule
 
 - & Dummy (MAD) & 0.858 & 0.959 & 1.787 & 0.267 & 1.844 & 53.762 & 27.455 & 10.538 & 0.508 & 65.886 & -\\

\midrule
 CIF & CGCNN (baseline) & 0.063 & 0.200 & 0.078 & 0.170 & 0.410 & 14.470 & 11.750 & 5.660 & 0.400 & 27.170 & 7.048\\
 
 \midrule

\multirow{6}{*}{Comp.} & Llama 2-7b-chat:0S & 40.722 & 86.989 & 91.114 & 53.109 & 2.005 & 125.545 & 73.335 & 71.208 & \colorbox[RGB]{255, 204, 229}{Inval.} & 365.525 & \colorbox[RGB]{255, 204, 229}{Inval.} \\

& Llama 2-7b-chat:5S & 0.968 & 87.324 & 90.542 & 0.207 & 1.883 & 61.121 & 27.672 & 23.093 & 0.598 & 57.394 & 0.704\\

& Mistral 7b-Instruct-v0.1:5S & 6.692 & 87.528 & 91.808 & 1.56 & 3.594 & 61.25 & 29.528 & 28.591 & 4.146 & 66.158 & 0.313\\

& Gemma 2-9b-it:5S & 5.736 & 87.117 & 90.989 & 0.445 & 1.894 & 42.779 & 24.141 & 15.256 & 1.949 & 56.098 & 0.49\\
\bottomrule
 
\end{tabular}
}
\label{tab:jarvis_dft_results}
\end{table}

\begin{table}
\centering
\caption{Results for SNUMAT. The $\downarrow$ and $\uparrow$ on the right of each property denote the MAE error and AUC score, respectively.}
\resizebox{1.0\textwidth}{!}{
\begin{tabular}{l l c  c  c  c  c  c  c | c  c }

\toprule

\textbf{Input} & \textbf{Model} & \textbf{Bandgap GGA $\downarrow$} & \textbf{Bandgap HSE $\downarrow$} & \textbf{Bandgap GGA optical $\downarrow$} & \textbf{Bandgap HSE optical $\downarrow$} & \textbf{Is direct $\uparrow$} & \textbf{Is direct HSE $\uparrow$} & \textbf{SOC $\uparrow$} & \textbf{Wtd. Avg} & \textbf{Wtd. Avg.}\\
& & & & & & & & & \bf (MAD:MAE) $\uparrow$ & \bf ROC AUC $\uparrow$\\

 \midrule

- & Dummy (MAD) & 0.875 & 1.092 & 1.154 & 1.343 & - & - & - & - & -\\

\midrule
 
 CIF & CGCNN (baseline) & 0.422 & 0.484 & 0.668 & 0.732 & 0.691 & 0.675 & 0.800 & 1.973 & 0.722\\
 
 \midrule

\multirow{6}{*}{Comp.} & Llama 2-7b-chat:0S & 1.099  & 1.152  & 0.998 & 1.564  & 0.503 & 0.484 & \colorbox[RGB]{255, 204, 229}{Inval.} & 0.940  & \colorbox[RGB]{255, 204, 229}{Inval.}\\

& Llama 2-7b-chat:5S & 0.691  & 0.823  & 1.339  & 1.144 & 0.475 & 0.468 & 0.455 & 1.157 & 0.466\\

& Mistral 7b-Instruct-v0.1:5S & 1.102 & 2.333 & 1.141 & 2.494 & 0.466 & 0.499 & 0.500 & 0.703 & 0.488\\

& Gemma 2-9b-it:5S & 1.182 & 0.935 & 1.876 & 1.585 & 0.530 & 0.485 & 0.719 & 0.843 & 0.584\\
\bottomrule
 
\end{tabular}}
\label{tab:snumat_results}
\end{table}

\begin{table}
\centering
\caption{Results for GNoME. The $\downarrow$ on the right of each property denotes the  MAE error.}
\resizebox{1.0\textwidth}{!}{
\begin{tabular}{l l c  c  c  c  c  c  | c  }

\toprule

\textbf{Input} & \textbf{Model} & \textbf{FEPA $\downarrow$} & \textbf{Bandgap $\downarrow$} & \textbf{DEPA $\downarrow$} & \textbf{Total energy $\downarrow$} & \textbf{Volume $\downarrow$} & \textbf{Density $\downarrow$} & \textbf{Wtd. Avg.}\\
& & & & & & & & \bf (MAD:MAE) $\uparrow$ \\

 \midrule

 - & Dummy (MAD) &  0.484 & 0.385 & 0.033 & 62.752 & 193.386 & 2.075 & -\\

 \midrule
 
 CIF & CGCNN (baseline) & 0.014 & 0.045 & 0.012 & 8.431 & 24.274 & 0.037 & 19.478 \\
 
 \midrule

\multirow{6}{*}{Comp.} & Llama 2-7b-chat:0S & 251.420 & 2.170 & 233.95 & 713.047 & 425.052 & 5.633 & 0.164 \\

& Llama 2-7b-chat:5S & 2.494 & 4.471 & 0.131 & 81.987 & 192.289 & 2.398 & 0.499\\

& Mistral 7b-Instruct-v0.1:5S & 2.526 & 1.591 & 1.494 & 94.448 & 225.382 & 11.129 & 0.336\\

& Gemma 2-9b-it:5S & 2.665 & 2.063 & 0.733 & 88.877 & 187.351 & 1.939 & 0.495\\
\bottomrule
 
\end{tabular}
}
\label{tab:gnome_results}
\end{table}

\begin{table}[hbt!]
\centering
\caption{Results for hMOF. The $\downarrow$ on the right of each property denotes the  MAE error.}
\resizebox{1.0\textwidth}{!}{
\begin{tabular}{l l c  c  c  c  c  c  c  | c  }

\toprule

\textbf{Input} & \textbf{Model} & \textbf{Max CO2 $\downarrow$} & \textbf{Min CO2 $\downarrow$} & \textbf{LCD $\downarrow$} & \textbf{PLD $\downarrow$} & \textbf{Void fraction $\downarrow$} & \textbf{Surface area m2g $\downarrow$}  & \textbf{Surface area m2cm3 $\downarrow$} & \textbf{Wtd. Avg.}\\
& & & & & & & & & \bf (MAD:MAE) $\uparrow$ \\

\midrule

 - & Dummy (MAD)& 2.245 & 0.11 & 3.506 & 3.59 & 0.181 & 1469.493 & 604.098 & - \\

 \midrule

 CIF & CGCNN (baseline) & 1.306 & 0.068 & 1.763 & 2.043 & 0.062 & 390.269 & 296.248 & 2.257\\
 
\midrule

 \multirow{6}{*}{Comp.} & Llama 2-7b-chat:0S & 199.955 & 60.032 & 381.040 & 447.976 & 0.361 & 3239.754 & 2593.456 & 0.174 \\
 
& Llama 2-7b-chat:5S & 3.305 & 1.902 & 3.696 & 3.500 & 0.191 & 2592.529  & 1651.846 & 0.655\\

& Mistral 7b-Instruct-v0.1:5S & 4.972 & 11.476 & 7.652 & 4.883 & 5.529 & 2590.894 & 1640.099 & 0.375\\

& Gemma 2-9b-it:5S & 3.307 & 2.238 & 3.613 & 3.691 & 0.236 & 2592.787 & 1659.706 & 0.624\\
\bottomrule
 
\end{tabular}
}
\label{tab:hmof_results}
\end{table}

\begin{table}[hbt!]
\centering
\caption{Results for Cantor HEA. The $\downarrow$ on the right of each property denotes the  MAE error.}
\resizebox{1.0\textwidth}{!}{
\begin{tabular}{l l c  c  c  c  | c  }

\toprule

\textbf{Input} & \textbf{Model} & \textbf{FEPA $\downarrow$} & \textbf{EPA $\downarrow$} & \textbf{Ehull $\downarrow$} & \textbf{VPA $\downarrow$}  & \textbf{Wtd. Avg. (MAD:MAE) $\uparrow$}\\

\midrule

 - & Dummy (MAD) &  0.108 & 0.792 & 0.175 & 0.470 & -\\

 \midrule

 CIF & CGCNN (baseline) & 0.012 & 0.016 & 0.018 & 0.164 & 17.780\\
 
 \midrule

\multirow{6}{*}{Comp.} & Llama 2-7b-chat:0S & 21.986 & 8.097 & 61.275 & 15.079 & 0.034\\

& Llama 2-7b-chat:5S & 0.121 & 1.205 & 0.188 & 0.477 & 0.867\\

& Mistral 7b-Instruct-v0.1:5S & 0.28 & 4.268 & 0.795 & 3.518 & 0.232\\

& Gemma 2-9b-it:5S & 0.732 & 1.219 & 0.181 & 1.754 & 0.508 \\

& MatBERT-109M & \textbf{0.033} & 0.049 & \textbf{0.034} & 0.189 & 6.769 \\
\bottomrule
 
\end{tabular}}
\label{tab:cantor_hea_results}
\end{table}

\begin{table}[hbt!]
\centering
\caption{Results for QMOF. The $\downarrow$ on the right of each property denotes the MAE error.}
\resizebox{1.0\textwidth}{!}{
\begin{tabular}{l l c   c   c   c  | c  c  }

\toprule

\textbf{Input} & \textbf{Model} & \textbf{Bandgap $\downarrow$} & \textbf{Total energy $\downarrow$} & \textbf{LCD $\downarrow$} & \textbf{PLD $\downarrow$}  & \textbf{Wtd. Avg. (MAD:MAE) $\uparrow$}\\

 \midrule
 
 - & Dummy (MAD) &  0.853 & 389.53 & 3.926 & 3.432 & -\\

 \midrule

 CIF & CGCNN (baseline) & 0.351 & 261.601 & 0.965 & 0.795 & 3.076\\
 
 \midrule

\multirow{6}{*}{Comp.} & Llama 2-7b-chat:0S & 0.947 & 1498.974 & 87.894 & 395.609 & 0.303 \\

& Llama 2-7b-chat:5S & 1.316 & 516.724 & 3.163 & 3.161 & 0.932 \\

& Mistral 7b-Instruct-v0.1:5S & 3.613 & 439.982 & 4.866 & 13.538 & 0.545\\

& Gemma 2-9b-it:5S & 0.81 & 567.827 & 4.302 & 3.276 & 0.925 \\

& MatBERT-109M & \textbf{0.468} & \textbf{229.838}  & \textbf{1.686} & \textbf{1.461}  & \textbf{2.049}  \\
\bottomrule
 
\end{tabular}
}
\label{tab:qmof_results}
\end{table}

\begin{table}[hbt!]
\centering
\caption{Results for JARVIS-QETB. The $\downarrow$ on the right of each property denotes the MAE error.}
\resizebox{1.0\textwidth}{!}{
\begin{tabular}{l l c  c  c  c  | c  }

\toprule

\textbf{Input} & \textbf{Model} & \textbf{FEPA $\downarrow$} & \textbf{EPA $\downarrow$} & \textbf{Total energy $\downarrow$} & \textbf{Indirect Bandgap $\downarrow$} & \textbf{Wtd. Avg. (MAD:MAE) $\uparrow$}\\

 \midrule
 
 - & Dummy (MAD) & 1.261 & 719.975 & 88.432 & 0.210 & - \\

 \midrule

 CIF & CGCNN (baseline) & 0.642 & 3.155 & 7.883 & 0.038 & 61.729\\
 
 \midrule

 \multirow{6}{*}{Comp.} & Llama 2-7b-chat:0S & 438.919 & 1949.724  & 514.448  & 1.000  & 0.188 \\
 
& Llama 2-7b-chat:5S & 1.553  & 694.574  & 85.674  & 0.161  & 1.047 \\

& Mistral 7b-Instruct-v0.1:5S & 1.819 & 819.901 & 145.431 & 0.534 & 0.643\\

& Gemma 2-9b-it:5S & 2.075 & 728.318 & 100.451 & 0.53 & 0.718 \\

& MatBERT-109M & 0.881  & 18.957  & 10.798 & 1.000  & 11.952 \\
\bottomrule
 
\end{tabular}
}
\label{tab:jarvis_qetb_results}
\end{table}

\begin{table}[hbt!]
\centering
\caption{Results for OQMD. The $\downarrow$ on the right of each property denotes the MAE error.}
\resizebox{1.0\textwidth}{!}{
\begin{tabular}{l l c  c  | c  }

\toprule

\textbf{Input} & \textbf{Model} & \textbf{FEPA $\downarrow$} & \textbf{Bandgap $\downarrow$} &  \textbf{Wtd. Avg. (MAD:MAE) $\uparrow$}\\
 
 \midrule
 
 - & Dummy (MAD) &  0.736 & 0.415 & -\\

 \midrule

 CIF & CGCNN (baseline) & 0.033 & 0.062 & 14.496 \\
 
 \midrule

  \multirow{6}{*}{Comp.} & Llama 2-7b-chat:0S & 38.194 & 2.166  & 0.105\\

& Llama 2-7b-chat:5S & 0.726  & 0.318  & 1.160 \\

& Mistral 7b-Instruct-v0.1:5S & 1.018 & 0.868 & 0.601\\

& Gemma 2-9b-it:5S & 1.079 & 0.678 & 0.647\\

& MatBERT-109M & 0.096 & \textbf{0.107} & 5.772 \\

& LLM-Prop-35M & \textbf{0.080}  & 0.146 & \textbf{6.020} \\ 

\midrule

\multirow{6}{*}{CIF} & Llama 2-7b-chat:0S & 80.815 & 8.767 & 0.028 \\

& Llama 2-7b-chat:5S & 0.700  & 0.240  & 1.391 \\

& Mistral 7b-Instruct-v0.1:5S & 2.963 & 1.852 & 0.236\\

& Gemma 2-9b-it:5S & 0.903 & 0.378 & 0.957\\
 
& MatBERT-109M & 0.053 & \colorbox[RGB]{204, 229, 255}{\textbf{0.058}} & 10.521 \\

& LLM-Prop-35M & \colorbox[RGB]{204, 229, 255}{\textbf{0.039}} & 0.129  & \colorbox[RGB]{204, 229, 255}{\textbf{11.041}}\\

\midrule

\multirow{4}{*}{Descr.} & Llama 2-7b-chat:0S & 29.677 & 2.221 & 0.106 \\

& Llama 2-7b-chat:5S & 0.742  & 0.283  & 1.230 \\

& Mistral 7b-Instruct-v0.1:5S & 1.247 & 0.693  & 0.595 \\

& Gemma 2-9b-it:5S &  0.821 & 0.441 & 0.919 \\

& MatBERT-109M & 0.049 & \bf 0.059  & \textbf{11.027} \\

& LLM-Prop-35M & \textbf{0.045} & 0.114  & 9.995 \\

\bottomrule
 
\end{tabular}
}
\label{tab:oqmd_results}
\end{table}

\begin{table}[hbt!]
\centering
\caption{Results for OMDB. The $\downarrow$ on the right of each property denotes the MAE error.}

\begin{tabular}{l l c | c}

\toprule

\textbf{Input} & \textbf{Model} & \textbf{Bandgap $\downarrow$} & \textbf{MAD:MAE $\uparrow$}\\

 \midrule
 
 - & Dummy (MAD) &  0.803 & -\\

 \midrule

 CIF & CGCNN (baseline) & 0.292 & 2.751\\
 
 \midrule

 \multirow{6}{*}{Comp.} & Llama 2-7b-chat:0S & 0.907 & 0.885\\
& Llama 2-7b-chat:5S & 0.796 & 1.009\\
& Mistral 7b-Instruct-v0.1:5S & 3.614 & 0.222 \\
& Gemma 2-9b-it:5S & 0.831 & 0.967 \\
& MatBERT-109M & \textbf{0.517} & \textbf{1.554}\\
& LLM-Prop-35M & 0.533 & 1.507 \\

\midrule

\multirow{4}{*}{CIF} & Llama 2-7b-chat:0S &  5.051 & 0.159\\
& Llama 2-7b-chat:5S & 0.864 & 0.930\\
& Mistral 7b-Instruct-v0.1:5S & \colorbox[RGB]{255, 204, 229}{Inval.} & \colorbox[RGB]{255, 204, 229}{Inval.}\\
& Gemma 2-9b-it:5S & \colorbox[RGB]{255, 204, 229}{Inval.} & \colorbox[RGB]{255, 204, 229}{Inval.}\\
& MatBERT-109M & \textbf{0.452} & \textbf{1.777} \\
& LLM-Prop-35M & \textbf{0.452} & \textbf{1.777} \\

\midrule

\multirow{4}{*}{Descr.} & Llama 2-7b-chat:0S & 5.194 & 0.155\\
& Llama 2-7b-chat:5S & 0.802 & 1.001 \\
& Mistral 7b-Instruct-v0.1:5S & \colorbox[RGB]{255, 204, 229}{Inval.} & \colorbox[RGB]{255, 204, 229}{Inval.}\\
& Gemma 2-9b-it:5S & 1.284 & 0.626\\
& MatBERT-109M & \colorbox[RGB]{204, 229, 255}{\textbf{0.435}} & \colorbox[RGB]{204, 229, 255}{\textbf{1.847}}\\
& LLM-Prop-35M & 0.485 & 1.656 \\

\bottomrule

\end{tabular}
\label{tab:omdb_results}
\end{table}

 \end{document}